**2017.06.12       36 pages and 7 figures**

**ESA has in April 2017 approved our proposal for study of a GaiaNIR mission. References and links are given on p.8 to our proposal and to a shorter version, both based on the following presentation and intense discussions led by David Hobbs as Principal Investigator.**

**Updates in red, e.g.**: p.13: Sect. 2.7 on "Wide Binaries" is now complete.
The previous version of this report from 2015.12.14 is placed as version v6 of
https://arxiv.org/abs/1408.2190 and contains information about previous updates.

# Absolute astrometry
# in the next 50 years

*Erik Høg, Niels Bohr Institute, Juliane Maries Vej 30, 2100 Copenhagen Ø, Denmark*
*Ehoeg@hotmail.dk*

ABSTRACT: With Gaia in orbit since December 2013 it is time to look at the future of fundamental astrometry and a time frame of 50 years is needed in this matter. A space mission with Gaia-like astrometric performance is required, but not necessarily a Gaia-like satellite. It should be studied whether this can be obtained within the budget of a medium-size ESA mission. A dozen science issues for a Gaia successor mission in twenty years, with launch about 2033, are presented and in this context also other possibilities for absolute astrometry with milliarcsecond (mas) or sub-mas accuracies are discussed. The three powerful techniques: VLBI, the MICADO camera on the E-ELT, and the LSST are described and documented by literature references and by an extensive correspondence with leading astronomers who readily responded with all the information I needed. In brief, the two Gaia-like missions would provide an astrometric foundation for all branches of astronomy from the solar system and stellar systems, including exo-planet systems, to compact galaxies, quasars and dark matter (DM) substructures by data which cannot be surpassed in the next 50 years.

# 1. Introduction

### 1.1  Overview

Please think where you and other astronomers need high-precision astrometry in the next 50 years. Think and let me know your science case! The main section **2. Science cases** on p.6 gives an overview of the presently proposed cases. Further *main sections* are: **3. Reference frames** on p.17, **4. Optical and radio astrometry** on p.18, **5. Densification and maintenance of reference frames** on p.23, **6. References** on p.27, **7.** Appendix A: **Astrometry with the MICADO camera** on p.31, and **8.** Appendix B: **Naming the Gaia successor mission** on p.34**.**

The provisional name "Gaia successor" is preferred in this report instead of "Gaia twin" as being more suited since there are important differences from Gaia in the proposed follow-up mission. The name *Gaia* (not Gaia-2 or Gaia 2) could be used if a specific code name is wanted in discussions. Another name for the mission has been proposed: *Rømer* or *Roemer* with reasons given in Appendix B.

A Gaia-like astrometric performance is required, not necessarily a Gaia-like satellite. It should be studied whether this can be obtained within the budget of a medium-size mission. The main issue is to build on the Gaia results of all-sky absolute astrometry for a billion stars. The Gaia successor should provide astrometry with equal or better accuracy for the same stars, and a common solution of the data from the two missions will give improved parallaxes and greatly improved proper motions. It would provide a new astrometric foundation for astrophysics, cf. Høg (2014d).

With the announcement in December 2014 of the new ESA launcher Ariane 6 as a successor to Ariane 5 the road to a larger more performing Gaia successor is opened and a launch with Ariane 6 will be much cheaper. This option for a Gaia successor, with better accuracy, limiting magnitude and resolution, may be called Gaia6. The performance and technological challenges are outlined in Høg (2015a), based on correspondence with Giuseppe Sarri, Gaia project manager. The expected performance is e.g. 10 microarcsec at 15 mag and a limit at 21 mag or fainter. But this should be elaborated taking into account the experiences with the flying Gaia, and the science cases should be described – I am inviting your contribution to both issues. This option shall not be further considered here.

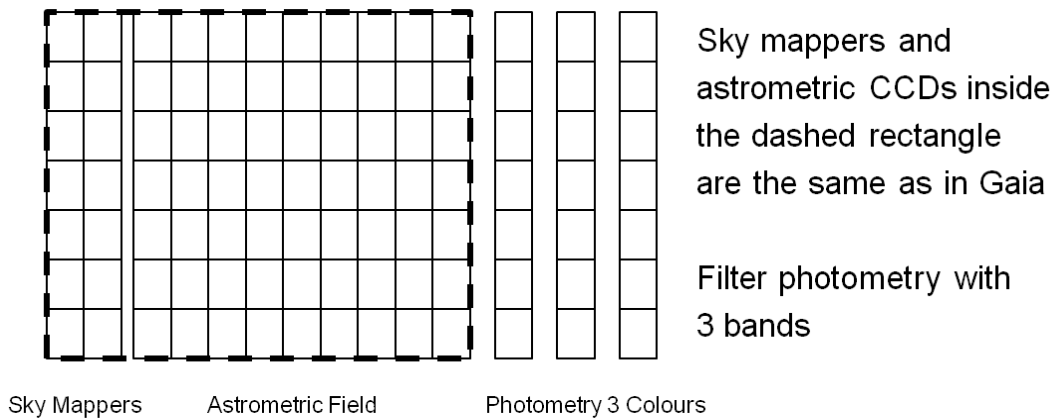

**Figure 1.** The proposed focal plane assembly. The filter photometry is required to enable correct chromatic corrections of the astrometric observations at every field crossing. The 3 filter bands shown here may well be augmented to 4 bands if wanted. Observations of bright stars to 3rd mag are obtained by way of special gates on the CCDs, see Gaia (2011: 2).

### 1.2 Design and basic performance

We assume here that the new mission would be launched 20 years after Gaia and have the same nominal duration of 5 years, but it is in July 2014 expected that Gaia has sufficient consumables for 7.5-9.5 years operation. If the new mission would also last for nearly 10 years and if it would be launched 30 years after the first, the errors on the resulting parallaxes would be further reduced with a factor 2.0 and the proper motions would have 20 times smaller errors than the "nominal Gaia", but the following discussion is based on 10 times smaller errors for proper motions.

With this mission goal, a payload design is proposed here. No other proposals for a Gaia successor are known, according to Anthony Brown in January 2015. There will be a brain storming session of a few days in Cambridge in the week of July 6.

The proposed focal plane configuration is shown in Figure 1 and is very similar to the focal plane of Gaia seen at Gaia (2011). As stated in my original proposal in May 2013, filter photometry is to be preferred to make chromatic corrections of the astrometric observations. This should replace the low-dispersion BP/RP photometry in Gaia because the filter photometry gives better angular resolution and the BP/RP photometry from Gaia will then already be available, but also because much accurate multi-colour photometry is becoming available for astrophysics. The medium-dispersion RVS instrument is hardly needed because the radial velocities and the spectra from the Gaia RVS will be available and results will come from many other surveys.

**Table 1.** Standard errors expected from Gaia before launch for a 5 year mission: in parallax ($\sigma_\pi$), position at mid-epoch ($\sigma_0$), and annual proper motion ($\sigma_\mu$), versus G magnitude for an unreddened G2V star (V-$I_C$ = 0.75 mag, V-G = 0.16 mag). Units are µas for position and parallax and µas yr$^{-1}$ for proper motion. The values are sky averages for a uniform stellar distribution. A figure and a table on the Gaia website show the variations across the sky. For stars brighter than G = 12 mag, the average performance is quoted over the range G = 6-12 mag. The last line contains the quantity DV in kpc, the distance at which the standard error of the tangential velocity from Gaia is 1 km/s, obtained as $DV=0.2110\ \sigma_\mu^{-1}$. The value of DV will be 10 times larger with results from two Gaia-like missions. – It is understood in Gaia (2014) from ESA on 22 July 2014 after the end of the Gaia commissioning phase that the pre-launch astrometric standard errors up to 14 mag will be obtained with a 20 % margin, but that a gradual increase of errors for fainter magnitudes is predicted. At G=20 mag errors of 540 µas for parallaxes, due to noise from unexpected stray light in the spacecraft. The mission should now be able to continue for more than 5 years, for a total of 7.5-9.5 years, in case funding for these activities can be found, and stars as bright as 3 mag will be measured. But for the present report we should apply the pre-launch errors in this table, for the sake of consistency.

| G | < 12 | 12 | 13 | 14 | 15 | 16 | 17 | 18 | 19 | 20 |
|---|---|---|---|---|---|---|---|---|---|---|
| $\sigma_0$ | 5.2 | 5.3 | 7.9 | 12.2 | 19.2 | 30.5 | 48.8 | 79.7 | 135 | 247 |
| $\sigma_\pi$ | 7.0 | 7.1 | 10.6 | 16.4 | 25.8 | 41.0 | 65.7 | 107 | 182 | 333 |
| $\sigma_\mu$ | 3.7 | 3.7 | 5.6 | 8.6 | 13.6 | 21.6 | 34.6 | 56.4 | 95.9 | 175 |
| DV | 57.0 | 57.0 | 37.7 | 24.5 | 15.5 | 9.77 | 6.10 | 3.74 | 2.20 | 1.21 |

The astrometric accuracies in Table 1 have been copied from the Gaia website before launch. For two Gaia-like missions at an interval of 20 years the accuracy of the proper motions is assumed to be 10 times smaller. The distance DV in kpc, the distance at which the standard error of the tangential velocity V is 1 km/s from Gaia, is given in the 5$^{th}$ line of Table 1, copied from Ibata (2013). DV shows at a glance whether the motion of a given object is accurate enough for a given scientific purpose, with the distance d of an object we have $V=d/DV$ km/s. - The errors of Gaia astrometry, photometry etc. as expected before launch have most recently been given by Luri et al. (2014).

### 1.3 Photometric distances and kinematics

Distances of stars are needed to derive tangential velocities from proper motions. Distances may be obtained from the parallaxes observed with Gaia, but they must be supplemented by photometric-spectroscopic distances for remote stars. The latter method will be much strengthened through calibration with accurate trigonometric distances from Gaia which will provide, e.g., better than 1.0 percent accuracy for 10 million stars, most of these will be dwarfs. For giants of luminosity class III, the sample within 2 kpc from the sun will contain stars brighter than V=11.5 if we assume M=0.0 mag to be typical. With $\sigma_\pi$=7.1 µas a relative accuracy better than 7.1/500 or 1.4 percent is then obtained for distances of giants. More precise estimates of the number of stars may be obtained from the population synthesis Galaxy star count model TRILEGAL (TRIdimensional modeL of thE GALaxy[1], Girardi et al. 2005, 2012) as described in Perryman et al. (2014) Sect.3.1.

Palmer (2014) has studied luminosity calibrations with Gaia data and developed a general method which can be used to determine absolute magnitudes of any kind of stars. In his thesis, Max Palmer has so far addressed RR-Lyrae stars and Cepheids, but others could be treated in the same way, according to this author. The question of calibration is discussed in Høg (2015b).

The expected photometry from Gaia is presented by Luri et al (2014). It appears from Fig.17 that the photometric errors at G=19 are about 3 mmag for G and 8 mmag for the low-dispersion red GRP, but much larger for the blue band GBP. Absolute magnitudes may therefore perhaps be obtained from the spectra of faint red stars. But the main source for accurate absolute magnitude of faint Gaia stars would probably be the many photometric surveys and the 4MOST spectra, after luminosity calibration with the Gaia parallaxes.

**Table 2.** The radial velocities from 4MOST compared to the tangential velocities from Gaia and from a Gaia successor. - Lines 2 and 3: Standard errors of radial velocities from 4MOST and of proper motions from Gaia as expected before launch for a 5 year mission. At the distance 2 DV, with DV from Table 1, the tangential velocity from Gaia proper motions obtains the standard error 2 km/s. M is the absolute magnitude at that distance, assuming no interstellar absorption. Corresponding values from two Gaia-like missions are given in the last lines as 20 DV and M-5.

| G *mag* | <12 | 13 | 14 | 15 | 16 | 17 | 18 | 19 | 20 |
|---|---|---|---|---|---|---|---|---|---|
| $\sigma_V$ *km/s* | 2.0 | 2.0 | 2.0 | 2.0 | 2.0 | 2.0 | 2.0 | 2.5 | 3.4 |
| $\sigma_\mu$ *µas yr$^{-1}$* | 3.7 | 5.6 | 8.6 | 13.6 | 21.6 | 34.6 | 56.4 | 95.9 | 175 |
| 2 DV *kpc* | 114 | 75 | 49 | 31 | ***19.5*** | 12.2 | 7.5 | 4.4 | 2.4 |
| M *mag* | -8.3 | -6.4 | -4.5 | -2.5 | ***-0.5*** | 1.6 | 3.6 | 5.8 | 8.1 |
| 20 DV *kpc* | 1140 | 750 | 490 | 310 | ***195*** | 122 | 75 | 44 | 24 |
| M-5 *mag* | -13.3 | -11.4 | -9.5 | -7.5 | ***-5.5*** | -3.4 | -1.4 | 0.8 | 3.1 |

Spectra for radial velocities and photometry for classification and distances of faint stars may be obtained in great number with the 4MOST facility of ESO (de Jong et al. 2012, Fig.1) which will be running spectroscopic Public Surveys on VISTA nearly full time from about 2020. Radial velocities are required to obtain space velocities for unambiguous dynamical studies. Radial velocities of ≤2 km/s accuracy of the faintest stars observed by Gaia may be obtained with 4MOST, and the following Table 2 compares the performance of Gaia and 4MOST.

Amina Helmi has given me the reason for the use of RGB stars in Helmi et al. (2011): *"There are plenty of RGB stars for every (old) stellar population and they are bright, so one can probe large distances, far out in the galactic halo. MSTO (main sequence turnoff) are more common, and hence even more useful but one*

---

[1] At http://stev.oapd.inaf.it/trilegal

*is restricted to remain in the inner 20-30 kpc down to V ~ 20 – 22 mag."* Only RGB stars with masses lower than 1.5 $M_{sun}$ live long enough to be observed and their number is high enough to be statistically significant. We will here for simplicity adopt tracers of M=–0.5 to represent the RGB stars. This absolute visual magnitude is 5 mag brighter than the present Sun and it is concluded from Fig. 25.11 of Maeder (2009) showing the solar evolution, that the Sun will stay brighter than M= -0.5 during about 0.1 Gyr. It should be noted that the G band used in the astrometric field includes more red light than the V band so that the red RGB stars will appear brighter than assumed here, Gaia (2014) gives G as function of V and V-I.

It will be crucial for the determination of photometric distances to these stars that their absolute magnitudes can be accurately calibrated. The following studies of photometric calibration in clusters are available: Bellazini et al. (2009), Valentini et al. (2004a, b), and Rizzi et al. (2007). More recent studies summarized in Høg (2015b) show that distances with 20% accuracy are now obtained for normal single stars, main sequence and giants. An accuracy of 10% will perhaps be common standard by 2025 after calibration with Gaia data and an accuracy of 1 or 2 % may even be obtained for giants and some other types of stars.

It appears, e.g. at G=16 mag, that Gaia will obtain 2 km/s at 20 kpc with tracer stars of M= -0.5, and two Gaia missions will get this accuracy at 195 kpc for supergiant tracers of absolute magnitude -5.5. As an example related to dark matter substructures, let us consider the measurement of tangential velocities for RGB stars in a volume out to 60 kpc from the Sun. This corresponds to Fig.2 in de Jong et al. (2012) which shows the spatial and radial velocity substructure distribution of red giant branch stars on the sky for a stellar halo formed in the Aquarius project of cosmological simulations from Helmi et al. (2011). This study is based on the model of Cooper et al. (2010) where stars are "painted" into dark matter substructures, and then as these substructures merge with the host Aquarius halo, the stars suffer tidal disruption contributing to a stellar halo.

At 60 kpc, corresponding to the limit in the Fig.2 of de Jong, an RGB star of absolute magnitude M= -0.5 has the apparent magnitude G=18.4 mag. The present Gaia mission will obtain an accuracy for that star of 20 km/s and two missions would obtain 2 km/s, i.e. 10 times smaller errrors if accurate distances are available. One Gaia mission will obtain 2 km/s at 20 kpc for such a tracer star, that is the same accuracy as from two mission at 3 times larger distance, i.e. from a 30 times larger volume, cf. the following Figure 2.

**1.4 The future in a wider perspective**

The issues and tasks of absolute astrometry in the next 50 years will be discussed in the following and some science cases for a Gaia successor are collected in the hope that this might inspire others to think of further cases. There is of course a long way to go before this effort can be complete, we recall that it took six years, 1993-99, to produce the study report for GAIA (ESA-SCI(2000)4), including the 100 pages with science cases compiled by the GAIA Science Advisory Group.

Even if a very strong collection of science cases can be defined for a Gaia successor, some problems must be overcome before a proposal has a chance to win. Some of these will be discussed elsewhere with proper documentation, here three hurdles very briefly: (1) The proposed Gaia successor is so similar to Gaia that it offers no technological challenge and therefore is believed by many colleagues be without a chance within ESA, but we require a *Gaia-like performance*, not necessarily a Gaia-like satellite. (2) A scientifically very interesting astrometric mission may perhaps be proposed with a technological interest, but after many years of study it would be realized that the difficulties are too big. With GAIA e.g. it took five years, 1993-98, before interferometry was dropped completely. With hindsight we easily proved that the proposed interferometric design was not a good idea for global astrometry (Høg 2014b). (3) Astrophysicists tend to prefer purely astrophysical missions, leaving astrometry behind and neglected, in spite of the fundamental astrophysical importance of astrometric data.

One of my colleagues noted in a correspondence that only few awards have been nominated by astrophysicists for modern astrometry, the global space astrometry, although the space missions are providing a new astrometric foundation of astrophysics. We agreed that these missions have been (also in his opinion) poorly recognised in general. Perhaps, he added, we see the projects only from the inside, so are not in the best position to judge from a wider perspective?

A wider perspective including physics and biology is contained in a correspondence from December 2013 when Gaia was launched. My colleague at the Niels Bohr Institute Nils O. Andersen (Professor in experimental atomic physics and former dean of the Faculty of Science at Copenhagen University) has allowed me to quote his answer when I mentioned my concerns for the future of astrometry: *"A lesson may perhaps be learnt from atomic physics. In spite of the shower of Nobel prizes these years in the area of cold atom physics, entanglement, Bose-Einstein condensation, etc., there is a clear understanding in "the scientific community" that we all depend crucially on the very reliable and precise input from more classical areas of atomic physics such as transition probabilities, oscillator strengths, lifetimes, etc. etc., including highly charged ions of astrophysical interest. This is indispensable for all the new stuff. Could one hope for a similar insight among astrophysicists?"*

He continued later: *"I also see a clear analogy to the situation in biology where many classic taxonomists for a long time have felt somewhat lost compared to their perhaps more successful colleagues in molecular biology. However, today the museum collections are considered a very valuable and unique resource, now under the general heading of "biodiversity", that turns out to be diminishing on our globe at an alarming rate. Consequently, the national natural history museums are now considered an invaluable part of our large scale research infrastructure, parallel to CERN, ESO, etc., with their collections being coordinated on a European scale."*

The experience from the approval processes of Hipparcos and Gaia (see Høg 2011, 2011b and 2013c) shows that a space astrometry mission will always meet great resistance. A Gaia successor probably cannot win in an open competition with astrophysical projects. **Fundamental Astrometry**, i.e. high-precision absolute global astrometry should therefore be given a special status in the science program.

This section shall end with quotations from Malcolm Longair and Freeman Dyson (Malcolm Longair, astrophysicist, astronomer Royal of Scotland in 1980, head of the Cavendish Laboratory from 1997 to 2005; Freeman John Dyson, theoretical physicist and mathematician, famous for his work in quantum electrodynamics, solid-state physics, astronomy and nuclear engineering). Longair (2001): *"The bedrock of astronomy remains the compilation of what is out there.... It is invidious to single out surveys which I find particularly impressive, but I make an exception for the Hipparcos astrometric satellite."* Dyson (1988) dedicated half a page to Hipparcos in his book where he wrote: *"Hipparcos is the first time since Sputnik in 1957 that a major new development in space science has come from outside the United States."*

# 2. Science cases

## 2.1 Overview

Optical imaging of radio sources with optical counterparts is considered. The maintenance of astrometric reference frames in the long term is vital for the astrophysical analysis of high-resolution images obtained in different wavelengths. The high angular resolution of future large optical telescopes imposes requirement of a milliarcsecond or less on the accuracy of optical astrometric reference frames, a requirement which can be satisfied by Gaia but only for some time. On a longer time scale a satisfactory reference frame can only be provided if a Gaia successor is launched in twenty years. This situation is discussed in the following, including a discussion of further science cases for a Gaia successor mission, e.g. the measurement of positions and proper motions by the E-ELT and the discovery of quasars solely from their zero proper motions. Measurement of absolute parallaxes and proper motions by reference to distant compact galaxies with the 42 m ELT telescope is described in detail and the accuracy is estimated; the ELT is now planned to have 39 m aperture and first light is expected in 2026. It is shown that a reference frame covering one half of the sky with many billions very faint stars could be produced from LSST observations and the Gaia catalog. The accuracies of such a frame would be about 1.0 mas for positions and parallaxes and 0.2 mas/yr for proper motions.

We focus on *absolute astrometry* defined as positions and proper motions of celestial objects in an inertial coordinate system the International Celestial Reference System (ICRS). Absolute parallaxes are included and are defined relative to infinitely distant objects. Absolute astrometry of positions, proper motions and parallaxes can be obtained by measurement of large angles in the sky with the astrometric satellites Hipparcos and Gaia. Absolute proper motions and parallaxes of stars can also be obtained by measurement of the very small angles relative to galaxies as planned for the MICADO camera on the E-ELT.

---

*The astrometric foundation of astrophysics*

**Top science**

*from a new Gaia-like mission vs. a single Gaia:*

- *Imaging of radio/optical sources etc.:*
  *Positions 50 years from now >20 times smaller errors*
- *Dynamics of Dark Matter etc. from stellar proper motions:*
  *Tangential velocities with 10 times smaller errors*
  *in 30 times larger volume*
- *Stellar distances in >3 times larger volume*
- *Exoplanets: Periods up to 40 years, vs. Gaia P<10 yrs*
- *Quasars solely by zero motions: 100 times cleaner sample*
- *Solar system: orbits, asteroid masses…*
- *Astrometry and photometry with 0."1 resolution*
- *Astrometric binaries. Common proper motion pairs. Etc. etc.*

**Figure 2.** A long-lasting astrometric foundation of astrophysics will be obtained by a new Gaia-like mission launched 20 years after the first. For example, in 2066, 50 years from now, the positions from the two missions will have 20 times smaller errors than from Gaia alone. With 10 times smaller errors on proper motions and tangential velocities, the volume covered with a certain accuracy for a given type of stellar tracer becomes 30 times larger, and even more than that because the long-term proper motions are less affected by motion in binaries. The possibility to perform astrometry with all sensors covering 400 nm to NIR belongs in this list thus penetrating obscured regions as molecular clouds and the Galactic centre.

## 2.2 A dozen science issues

Two science cases for a Gaia successor mission are described elsewhere: Galactic dynamics by Ibata (2013) and long-period exoplanets by Høg (2013b) and in Section 2.6. Here follow further cases, and an overview of the top science is shown in Figure 2.

The science case for discovery and mapping of optical 2D-structure in radio sources is introduced. Use of the E-ELT with 42 m aperture for the purpose is discussed. The GMT with 25.7 m aperture is also mentioned because it may offer a larger field of view. The TMT with 30 m aperture will have a smaller field of 30 arcsec diameter.

The science case is introduced for a reference system to serve astrometry with the E-ELT which can lead to proper motions in clusters or dense areas with a precision of 2-6 µas/yr corresponding to about 1-3 km/s at 100 kpc distance.

A whole range of science cases is opened by the proposed high-resolution photometry, see the arrangement of filters in Figure 1 and the brief description below. Angular resolution of the filter photometry will be 0.14 arcsec (FWHM), ten times smaller than in the Gaia photometry with low-dispersion spectra.

The science case is briefly described in a following section for the detection of QSOs and possibly other kinds of extragalactic point sources solely from zero proper motion and parallax, unbiased by any assumptions on spectra. This issue has been studied in order to see how well this can be done with the smaller proper motion errors from two missions.

Science cases for detection and mapping of dark matter are briefly described in a letter from Jean Kovalevsky placed in a following section, including comments and references. Dark matter in the halo and disk of our galaxy, in globular clusters and outside of our galaxy are mentioned, ie at least four science cases are described.

The detection of astrometric binaries will reveal unseen companions which may in principle be low-luminosity objects of many kinds, e.g., exoplanets, brown dwarfs, neutron stars, black holes, especially with long period of 3-100 years which are difficult for spectroscopic and imaging investigation.

Astrometric binaries and common proper motion pairs hold clues to stellar formation and evolution and they can only be detected by astrometry, not by eclipsing or radial velocity measurements. About one half of all stars belongs to these kinds of systems – about one fourth belonging to astrometric binaries and the same fraction to common proper motion pairs.

The solar system is discussed with overviews of the future need for astrometry. It is argued that the studies need a reference frame to G=20 mag with an accuracy of 0.5 mas and that this can be provided in the future by a Gaia successor, but not by the proposed all-sky survey from the ground.

A focal plane design with all sensors covering 400 nm to NIR has been proposed by Høg, Knude & Mora (2015) resulting in a better penetration for astrometry in obscured regions as the Galactic centre and molecular clouds. Since November 2015 Timo Prusti is considering what ESA can do and I am corresponding with several manufacturers of sensors about the technical aspects. Our further studies are summarized in Høg (2017).

In April 2017 ESA selected our proposal Hobbs et al. (2016) for study of a detector with NIR sensitivity for a Gaia successor mission, called GaiaNIR. A shorter version of our proposal is available as Hobbs & Høg (2017). ESA received 26 proposals for studies of new science ideas and selected 3 key themes, one of them was ours, see ESA (2017), where it is called "high-accuracy astrometry".

## 2.3 Photometry with high angular resolution

A whole range of science cases is opened by the proposed high-resolution filter photometry. This should be elaborated, but is here mentioned only briefly. The angular resolution along scan in the astrometric field is given by the typical 1-sigma value of the LSF of 1 pixel along scan, i.e 59 mas. So the typical FWHM is 2.35 pixel = 140 mas and very similar for the proposed filter photometry. In Gaia however the low-dispersion BP/RP spectra both have a typical length of 2650 mas at 15 mag, according to Gaia (2014) and information from Jos de Bruijne. The chromatic correction of astrometry for asymmetric objects, e.g. double stars with separation below 2 arcsec, compact galaxies, quasars will be much better for observations by the Gaia successor than by Gaia. Parallaxes and proper motions will be better for both missions after a new reduction of Gaia data using photometry from the successor. Lennart Lindegren wrote in May 2013: *"The argument for high angular resolution photometry is compelling, for it is certainly one of the weak points of Gaia."* see Høg (2013a, p.2). Now he commented: *"I do not have much to add to the case for filter photometry, except that a certain class of close binaries (with a colour difference between the components) could be detected using the colour-dependent position of the centroid. It could also help to identify more complex structures, e.g. in quasars."*

### 2.4 Astrometric detection of QSOs

Astrometric Detection of QSOs, i.e. solely from zero proper motion and parallax, unbiased by any assumptions on spectra, might lead to discovery of a new kind of extragalactic point sources. This issue has been studied by Fynbo & Høg (2014) and Heintz, Fynbo & Høg (2015) in order to see how well this can be done with the expected Gaia data and with the smaller proper motion errors from two missions.

Bailer-Jones et al. (2008) have developed a method using Gaia photometry by low-dispersion spectra and shown with simulated data, that it is possible to achieve a pure sample of quasars (upper limit on contamination of 1 in 40000) with a completeness of 65 per cent at magnitudes of $G = 18.5$, and 50 per cent at $G = 20.0$, even when quasars have a frequency of only 1 in every 2000 objects. This is a satisfactory sample of quasars for the inertial tie of the rotation of the Gaia coordinates. Our aim is different, to discover astrophysically interesting quasars in the remaining sample.

The incompleteness of quasar samples based on selection by optical photometry has been studied intensively for many years and it is now well established that such samples miss a substantial number of in particular dust-reddened quasars (see, e.g., Fynbo et al. 2013 for a recent study). Our strategy is to select quasar candidates solely on the basis of their lack of proper motion. This selection strategy also has the potential to select other extragalactic point sources, e.g. potentially new classes of objects. In order to examine the feasibility of this approach we needed to determine the number of false positives, e.g. how many stars will be selected in this way and where on the sky (or towards which galactic coordinates) will the problem of stellar contamination be most severe.

We have used a catalog generated for the Gaia mission (the so called GUMS data, the "Gaia Universe Model Snapshot"). This has enabled us to derive precise numbers for the expected true detections and for the false detections due to stars which happen to show zero motion. Fynbo & Høg (2014) predicted that two missions would be 100 times better than one Gaia mission since the number of false detections will be 100 times smaller. This follows because the proper motion errors will be 10 times smaller in both coordinates. This prediction from August 2014 has been confirmed in a paper by Heintz, Fynbo & Høg (2015) which has been published in A&A. For regions above 30 degrees latitude the ratio of QSOs to apparently stationary stars is above 50% and towards the poles about 80% when using Gaia data. With a Gaia successor in 20 years the ratio would improve dramatically at all latitudes.

### 2.5 Dark matter

*Jean Kovalevsky* wrote on 2014.04.28: *"Thank you for sending me this remarkable overview of problems and solutions for astrometry in the next 50 years. I enjoyed it very much. As you know, I have now been away from astrometry for several years (I try to understand the present advances in cosmology). So I am no more aware of the latest thoughts in astrometry, This is why your so well documented paper is a wonderful occasion for me to think a little bit about the fields in which astrometry can be a useful tool.*

*There is one field your paper does not mention: "dark matter". … Anyhow, congratulations for your work."*

Dark matter was treated soon after enjoying the help by Mattia Vaccari (SKA SA Post Doctoral Fellow, Physics Department, University of the Western Cape, Cape Town, South Africa) with the following, and all this less than two weeks after I asked his comment to the dark matter issues.

Inside our Galaxy, the results from two Gaia-like missions as discussed by Ibata (2013) may be characterized by about 1.0 km/s accuracy of the tangential velocity for giants or horizontal branch stars at a distance of 40 kpc, thus allowing study of internal motions in nearly all clusters of our Galaxy, including the outer halo globular clusters. This accuracy of tangential velocities is derived from Table 1 which predicts 1.0 km/s by two missions for a star of $M_V$=0 mag at the distance 37.4 kpc where the star would obtain the apparent magnitude 18.0. At 20 kpc where m-M=16.5 mag the accuracy for these stars will be 2.6 km/s with Gaia and 0.26 km/s with two missions.

In Moni Bidin et al. (2010) the dynamical surface mass density ($\Sigma$) is estimated at the solar Galactocentric distance between 2 and 4 kpc from the Galactic plane, as inferred from the observed kinematics of the thick disk. The authors find only slight indications of dark matter. They believe that successfully predicting the stellar thick disk properties and a dark disk in agreement with their observations could be a challenging theoretical task.

Detection of dark matter substructures to a distance of 60 kpc in the Galactic Halo according to Helmi et al. (2009) was described as an example in the preceding introduction.

The unexpected fast rotation of the outer parts of galactic disc has long been known. The shape of the rotation curve in the outer Galaxy, is crucial in determining the gravitational field and its associated putative dark matter distribution as discussed by Famaey (2012). The paper states that the case for the presence or absence of the various observed or proposed "features in the outer Galaxy rotation curve will certainly be settled with Gaia data, and will be of prime importance for ascertaining the presence or absence of dark matter substructures inside the Milky Way disk, in parallel with a better understanding of the precise density distribution of stars in the outer Galaxy, including features such as the Monoceros overdensity."

The outer part of the Galaxy is also subject of a paper by Feast et al. (2014) which reads as follows: "It is instructive to examine why the outer regions of a galactic disk flare. In the inner parts of a galactic disk the gravitational force k(z) at height z perpendicular to the galactic plane is dominated by the strong concentration of stars there. As we move to greater galactocentric radii, however, the concentration of stars drops dramatically, k(z) decreases and is increasingly dominated by the effects of dark matter… It is highly desirable that the gravitational field in the outer Galaxy be investigated using young stars for which good distance estimates can be made. Classical Cepheid variables are by far the best stars for this purpose."

The globular NGC 2419 at the very large distance of 87 kpc was found by Ibata et al. (2012) to contain dark matter in an amount of probably less than 6% of the luminous mass inside the tidal limit. But the authors add at the end: " … the presence of a dark matter halo around NGC 2419 cannot be fully ruled out at present, yet any dark matter within the 10 arcmin visible extent of the cluster must be highly concentrated and cannot exceed $1.1 \times 10^6$ Solar masses (99% confidence), in stark contrast to expectations for a plausible progenitor halo of this structure."

Turning now to dark matter outside our Galaxy, a paper by Feldmann & Spolyar (2013), says that Cold Dark Matter (CDM) theory predicts the existence of a large number of starless dark matter halos surrounding the Milky Way (MW). However, clear observational evidence of these "dark" substructures remains elusive. A detection method is presented based on the small, but detectable, velocity changes that an orbiting substructure imposes on the stars in the MW disk. Using high-resolution numerical simulations the authors estimate that Gaia should detect the kinematic signatures of a few starless substructures provided the CDM paradigm holds. Such a measurement will provide unprecedented constraints on the primordial matter power spectrum at low-mass scales and offer a new handle onto the particle physics properties of dark matter.

The detection of dark matter outside the Galaxy by two Gaia-like missions launched at an interval of 20 years was discussed in Ibata (2013) where it is concluded that the two missions would unveil the details of the dark matter distribution around the Milky Way, and they will enable us to finally uncover the distribution of dark matter in nearby dwarf spheroidal galaxies, which are thought to be the remnants of the cosmological building blocks that merged to form large galaxies.

In summary, a combined use of more accurate radial and tangential velocities and of the stellar content of the Galaxy to be obtained with Gaia and future missions would shed much more light on these questions for globular clusters, the Galaxy disk and halo and the Milky Way neighbourhood. The specific gain for the investigation of dark matter by a Gaia successor mission should be further studied. A large gain from the ten times improvement of proper motions would be expected because the only known manifestation of DM is that it drives the motions in the universe through its gravitational force. One of the most powerful methods to detect DM therefore arguably hinges on the astrometry of the two-dimensional proper motions, besides of the only one-dimensional radial velocities.

Later on, I received from Jesús Zavala Franco (Postdoc at Dark Cosmology Centre, Niels Bohr Institute): Current observations of the stars in the MW dwarf spheroidals (dSphs) provide only an incomplete picture of their full 6D phase-space distribution. The information available is only three-dimensional: the 2D projected spatial distribution and the 1D radial velocity. This creates the well known mass-velocity-anisotropy degeneracy, which prevents a reconstruction of the DM distribution in the dSphs. Kinematical information in the tangential direction provided by space astrometry would break this degeneracy.

Jesús Zavala Franco also mentioned two recent papers related to dark matter. (1) Daylan et al. (2014) reconsider the gamma ray signal from the region around the Galactic Center which is consistent with the emission expected from annihilating dark matter. They find that the signal is observed to extend to at least 10 degrees from the Galactic Center. A precise measurement of the space velocities of stars in the inner Galactic region (within 10 degrees) from Gaia and a successor, would greatly constrain the dark matter distribution in this region. This would allow us to establish if it is possible to have a steep dark matter density profile as is required by the hypothetical gamma-ray signal. For such a study, it is noted that tangential velocities for stars of 20 mag at a distance of 8 kpc will obtain an accuracy of 7 km/s from Gaia according to Table 1 and 0.7 km/s from two missions, using tracers of respectively M=5.5 and 0.5 mag if no absorption is assumed. It is recalled that Gaia and also the successor will measure in a band around the visual where the effect of interstellar absorption is large at low latitudes. (2) Bovy & Rix (2013) present and apply rigorous dynamical modeling with which they infer unprecedented constraints on the stellar and dark matter mass distribution within our Milky Way, based on large sets of phase-space data on individual stars. The authors use especially a set of 23,767 G-dwarfs with well-determined measurements, i.e. SSDS photometry, including ([Fe/H],[α/Fe]), distances, proper motions and radial velocities. It would be interesting to know the improvement that could be expected when Gaia data become available, and what further improvement could be expected from a Gaia successor mission.

## 2.6 Astrometric binaries

An astrometric binary is detected if the star has a motion on the sky indicating the gravitational effect of an unseen companion, i.e. if the motion cannot be represented by the standard five-parameter astrometric solution giving position, proper motion and parallax. Unseen companions may in principle be low-luminosity objects of many kinds, e.g., exoplanets, brown dwarfs, neutron stars, black holes.

The treatment of binaries in the Gaia data reduction is briefly outlined in a report by Høg (2014e). It is expected from simulations that about 59% of systems in the Gaia survey to G<20 mag will be single stars, and that some 60 % of all stars will belong to a binary or ternary system. About 80% of non-single systems would have periods above 5 years. Detection of binarity is possible by determination of orbits and sometimes only by an observed acceleration, a total of 6 million such detections are expected from simulations before launch among the one billion stars detected by Gaia, i.e. for about 0.6 % of these stars. - Detection of binarity from a combination of Gaia, Hipparcos and Tycho-2 data is briefly reviewed, incited by recent simulations at the Lund Observatory. Binaries and exoplanets with long periods could be detected among millions of stars from two Gaia-like missions, see Figure 3. If results from the Hipparcos mission are included, orbits for, e.g. Saturn-Sun like systems would be well determined if the Hipparcos accuracy is sufficient. Exoplanet detection has also been discussed by Høg (2013b).

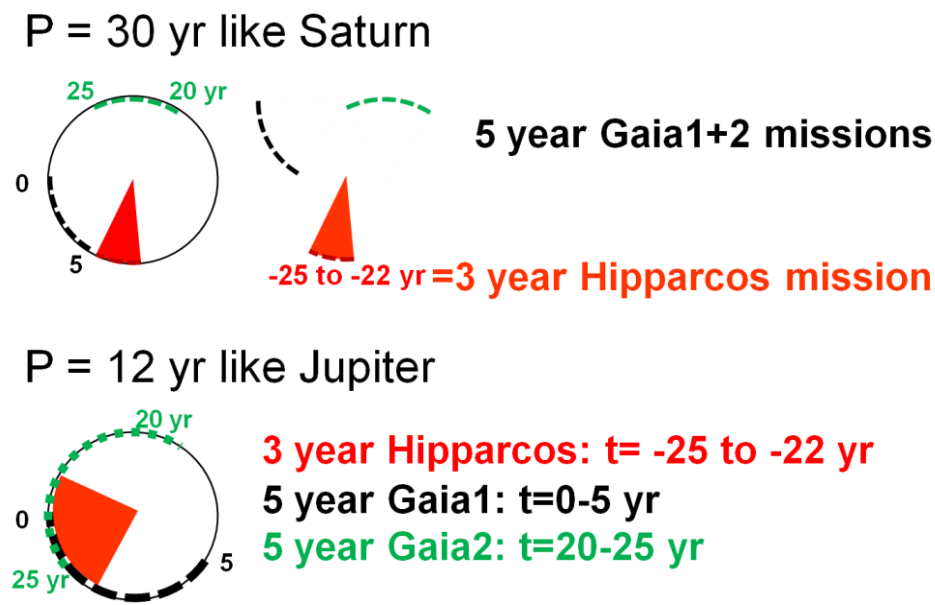

**Figure 3.** Data from Hipparcos and two Gaia-like missions may be used for orbit determination of exoplanets, here shown for the examples of Saturn- and Jupiter-like cases. Circular and face-on orbits are shown for simplicity, but it appears that 12 year orbits will be well covered for all combinations of orbital elements by two Gaia-like missions only. For 30 year orbits, however, the inclusion of Hipparcos data will be important if the accuracy is sufficient.

Black holes as supernova remnants from heavy stars are found as X-ray transients with short periods. A black hole as remnant in a binary of long period will usually escape from its companion because the gravitational attraction becomes too weak to keep after the mass-loss during the explosion so that the two bodies go into hyperbolic orbits. But this does perhaps not always happen in case the explosion is asymmetric. The black hole might receive an impulse in such a direction that the orbital velocity is decreased just sufficient to keep the pair together. The presence of a black hole would appear from a large total mass if an orbit can be determined.

Gould & Salim (2002) have studied the possibility that some fraction of luminous stars ended as black hole remnants, without producing a supernova (a ``failed'' supernova). They show that, under plausible assumptions, the Hipparcos catalog could contain a number of astrometric binaries with black hole companions. No black hole astrometric binaries are found in Hipparcos catalog, but the main uncertainty in this estimate is the binary companion mass function, which in itself is not well constrained by Hipparcos.

The authors show that using future space-based astrometric missions, an accurate measurement of the rate of supernovae that fail could be obtained by finding black hole binaries and measuring the progenitor mass function.

Astrometric binaries have been discussed by Makarov & Kaplan (2005) based on the results of the Hipparcos mission and older ground-based astrometric catalogs as contained in the Hipparcos and the Tycho-2 Catalogues. The authors present two catalogs of several thousand astrometric binaries as being useful in the ongoing quest for low-mass binaries and brown dwarfs in the solar neighborhood. The method of astrometric motion analysis is sensitive in the most difficult area of orbital parameters for both spectroscopic and imaging investigation, i.e., orbital periods between 3 and 100 yr. Makarov & Kaplan note that astrometric binaries may prove to be a considerable difficulty in the complicated data reduction systems for Gaia. If, for example, brown dwarf companions are widespread in long-period binaries, a large number of reference stars may be completely unsuitable for processing with the standard astrometric model of linear motion and parallax.

Statistically, about 45-50% of all field stars are binary or multiple, according to Valeri Makarov (Chief, Space Astrometry & Instrumentation Division, United States Naval Observatory). Perhaps half of them have orbital periods longer than 5 years. So a quarter of all stars will be affected by the difference between long- and short-term proper motion. These predictions agree with those in Høg (2014e).

### 2.7 Wide binaries

We now prefer the term Wide Binaries instead of common proper motion pairs (CPM) following an advice from Jeff Andrews in Høg & Andrews (2017). This report of 3 pages contains discussion of this matter including a result from the use of TGAS to detect wide binaries. - **Very short abstract:** The components of wide binaries are expected to form with the same metallicity at the same time and are evolving independently which makes them extremely valuable for stellar evolution. Wide binaries will be detected by common proper motions of the components and the samples collected by two Gaia-like missions will be about 100 times cleaner than from one mission.

Previous text of this paragraph: Astrometric binaries and common proper motion pairs (CPM) hold clues to stellar formation and evolution and they can only be detected by astrometry, not by eclipsing or radial velocity measurements. About one half of all stars belongs to these kinds of systems – about one fourth belonging to astrometric binaries and the same fraction to common proper motion pairs, according to Christian Westhues (Ruhr University, Bochum).

The paper by Chini et al. (2014) of common proper motion pairs shows that our census of the nearby CPM companions is far from complete, even within our closest neighbourhood of 25 pc. The authors emphasize that it is of principal importance to know the local inventory as this has far-reaching consequences on all scales of astrophysical research. The locally existing degenerate stars mostly contribute to issues of the mass density, the stellar populations, and the early Galactic epochs. Simultaneously, there is a major interest in the star formation process in general because evidence is growing that stars often are created as multiple systems. On the other hand, the formation of very wide binaries is difficult to understand, because their observed separation can exceed the typical size of a collapsing cloudcore.

Gaia will greatly contribute to the inventory of CPMs and also a Gaia successor will do so - as should be quantified.

### 2.8 Solar system and small-field astrometry

Astrometric issues for solar system studies were discussed during July to October 2014 in a correspondence with a number of colleagues and this has been collected in Høg & Kaplan (2014) of which here follows a summary.

Tanga (2014) defines three levels of astrometric accuracy in order to show the increasing amount of science for the solar system obtained by better accuracy and he says that 100 muas would be useful for four specific scientific purposes. Tanga expects that 0.1 mas will only be obtained with observations from space and that 1 mas will be possible from the ground. Michael Shao (2014) comments on the report by Paolo Tanga and disagrees on some points which could, regrettably, not be resolved in the correspondence.

A new reduction of old astrometric observations of solar system objects will according to Arlot (2014) be obtained when the Gaia reference star catalogue will be available. Its accuracy will give an increase of the accuracy of the many old observations obtained since photography was introduced about 1890. A Gaia successor will secure high-precision astrometry in the solar system also in the far future and it appears that a measurement accuracy of 1 mas will be sufficient because the irregularity of the figure of the objects will set a limit. But Tanga (2014) claims that 0.1 mas would be useful.

Overviews of issues for *future* solar system studies from two colleagues in the USA, George Kaplan (US Naval Observatory, Washington, retired) and Hugh Harris (US Naval Observatory, Flagstaff) are given here *in extenso*: Kaplan (2014) and Harris (2014) and are also for the most part available in Høg & Kaplan (2014). These overviews give references and cover, with some overlap, all aspects of the solar system where astrometry is important: orbits of planets, moons, asteroids and NEOs, masses of asteroids, occultations of asteroids and KBOs, and families of asteroids and KBOs. The roles of astrometry from the ground, from Gaia and from a Gaia successor are discussed by both, but *not* small-field astrometry from space.

Harris expects accuracies of 1-5 mas for ground-based observations with small-field astrometry when Gaia results become available in the form of a very accurate absolute reference frame with a large number of stars, close to one billion. The ground-based observations determine positions relative to the reference frame for other objects in the field, i.e. stars or solar system objects. For stars, the proper motions and parallaxes can be derived after years of observations. For solar system objects, orbits can be determined.

Consequently the same accuracies of 1-5 mas are expected for predictions of positions with the new orbits, representing improvement by a factor of 10-100 over the present. Harris and his colleagues now use a field size of typically 10 arcmin (i.e. a square with these sides), and reach standard errors of 3 mas for a single exposure, 2 mas if the reference field is 6 arcmin. He expects they will move toward smaller field sizes in the future.

A comparison with the expression by Lindegren (1980) for the astrometric errors due to the atmosphere is given at the end of Harris (2014). This expression indicates that the improvement with smaller fields goes with the field to the power 0.25, much more slowly than suggested by the numbers given by Harris. This issue should be further investigated, but on the basis of this finding and these reports I suggest that an accuracy of 1 mas is the best possible from the ground when using ordinary telescopes, i.e. without wave-front correctors.

If 1 mas is the limit, an accuracy (random error) of 0.5 mas is adequate for the reference stars since such a frame would contribute only 0.1-0.2 mas to the standard error of the object. It should be noted that the systematic errors with astrometry satellites are much smaller than the random errors, thus for Gaia systematic errors about 0.001 mas is expected. With ground-based astrometry systematic errors are often comparable to the random errors.

The reference frame should contain all stars to G=20 mag, but it need not go fainter for the sake of solar system work, as explained in Harris (2014). The limit of 1 mas is expected for observation of a star or solar system object in a reference frame *in a single night*. For stars, the accuracy can be improved by observations on many nights if the reference system is more accurate. But this is not required for a solar system object, since e.g. a KBO at a distance of 40 AU will move about 1.5 arcmin per day due to the Earth's motion and therefore soon appear among other reference stars.

The Gaia frame will have errors of 1.8 mas at G=20 mag in 2026, 3.5 mas in 2036, and 8.8 mas in 2066 as explained below. This could not at all satisfy solar system observers. It has been suggested to use the more accurate Gaia stars of G=16 and brighter, but they are too sparse to fill the small fields required to obtain the 1 mas accuracy in observations.

A solution to this problem would be a Gaia successor as advocated here. Another solution to the problem would be a densification of the optical reference frame as proposed by Zacharias in section 5. An all-sky survey with 1-meter class telescopes could give a frame with 2 mas accuracy of stars to G=22 mag which is however not the accuracy wanted for solar system work. Thus, a Gaia successor providing 0.5 mas accuracy or better at G=20 mag is required.

### 2.9 Spacecraft navigation and attitude control

The importance of two Gaia-like missions for spacecraft control should be considered. The missions can provide a long-term astrometric foundation of spacecraft navigation and attitude control. John Leif Jørgensen (Professor and leader of the division for measurement and instrumentation at the Danish Technical University) says that a better foundation is required for many, in particular high accuracy formation flights proposed for the future. Perhaps the reference frame to be provided by Gaia will be sufficient for a long time. Astrometrists should care about these aspects, not only about the astrometric needs within astronomy.

# 3. Reference frames

### 3.1 Celestial reference frames

The position of a celestial object should be given in the International Celestial Reference System (ICRS), see e.g. USNO (2013) from where the following about reference frames is quoted. The ICRS is realized by catalogs, called frames, of reference objects with positions at the actual wavelength of observation, e.g., radio or optical. The *International Celestial Reference Frame (ICRF or ICRF1)* is a catalog of adopted positions of 608 extragalactic radio sources observed with VLBI, all strong (greater than 0.1 Jy) at S and X bands (wavelengths 13 and 3.6 cm). Most have faint optical counterparts (typically with visual magnitudes fainter than 18) and the majority are quasars. The current radio reference frame is ICRF2 with over 3400 sources (Fey et al. IERS Technical Note 35, 2009).

The ICRS is realized at optical wavelengths by stars in the Hipparcos Catalogue of 118,218 stars, some as faint as visual magnitude 12. Only stars with uncomplicated and well-determined proper motions (e.g., no known binaries) are used for the ICRS realization. This subset, referred to as the *Hipparcos Celestial Reference Frame (HCRF)*, comprises 85% of the stars in the Hipparcos catalog. Extensions to fainter magnitudes and to the infrared are detailed in USNO (2013). The past and present of reference frames is presented on the USNO website, but not the future which will be discussed here.

A few papers shall be mentioned with relation to the proposed use of reference frames at imaging of optical and radio sources. Present day faint reference stars (V~16) have typical accidental errors of 30mas but systematic errors up to 100mas can occur, according to Makarov et al. (2012). The same paper shows an example of a nearby and bright quasar. It appears as a bona fide reference quasar in the radio but is a horrible source in the visible light extending to at least 10 arcsec. Even at 100 times larger distance, further away than any known quasar, it would be marginally disturbed for a mission like Gaia. This illustrates the importance of using very good reference frames when comparing and superposing the images of quasars, as pointed out in the following.

N. Zacharias has informed me that systematic errors in the current optical reference stars to V = 16 (i.e. UCAC4) are *not* as large as 100 mas; more like 20 mas; remaining systematic errors of epoch 1900 Astrographic Catalog photographic plate positions are about 100 mas, also real centroid offsets between radio core and optical counterpart of QSOs can be over 100 mas

Tytler (1997) discusses the cosmology which becomes possible with the Very Large Telescope Interferometer (VLTI) using a space based astrometric reference frame. Position(?), parallax and proper motion from space and on a global reference frame for at least one reference star (V ≃ 16) in the isoplanatic patch of all interesting targets are required. The VLTI then gives positions to about 14 μas when the reference star is 10 arcsec away, and 30 μas at 30 arcsec.

Camargo et al. (2011) have investigated the differences between positions, as determined by optical (direct imaging) and Very Long Baseline Interferometry (VLBI) techniques, of extragalactic sources listed in the second realization of the International Celestial Reference Frame (ICRF2). The differences may reach more than 80 milliarcseconds and, taking into consideration that they are hardly explained only by statistical fluctuations or systematic errors in the optical reference frame used here, the authors argue that these differences can be related to the sources' X-band structure index (8.4 GHz). They conclude that the presence of the intrinsic structure should be taken into consideration when comparing the optical and VLBI positions of ICRF2 sources in the future. But N. Zacharias notes that the optical structure unfortunately can not be taken into account because it is below the resolution of optical data.

The study by Bourda et al. (2012) shows that less than half of the quasars initially selected for a VLBI reference frame were point-like, the others showing a significant structure above the 1 mas scale.

The future of VLBI holds much promise for pushing the accuracy of fundamental astrometric Reference Systems to their physical limits, and an overview is given by van Langevelde (2012).

### 3.2 Overview of accuracies from space missions

The expected standard errors of the Gaia catalog at G=20mag are 0.175mas/yr for proper motions and 0.247mas for positions at the mid epoch, presumably 2016. This implies position errors of 1.76mas in 2026, 3.5mas in 2036, and 8.8mas in 2066. The systematic error of Gaia positions is of the order 0.001mas so that the error of the average of many Gaia stars can be expected to decrease with the square root of the number of stars.

Parallaxes of all stars are given in all reference frames derived from the space missions and have to be applied to obtain the positions at a given epoch. The radial velocity is also given for the rare cases where it has to be applied.

The construction and accuracy of the Gaia Celestial Reference Frame (GCRF) have been described by Mignard (2008 and 2011). The frame should be inertiality as good as 0.5 μas yr$^{-1}$ as illustrated in Fig. 3 of Mignard (2011) which is 1000 times better than the Hipparcos frame.

Hipparcos has mid-epoch of observations J1991.5. Reference system ICRS, coincidence with ICRS with 0.6mas errors on all 3 axes, deviation from inertial 0.25mas/yr on all 3 axes.

Tycho-2 has precisely the same accuracy of its frame as given for Hipparcos. N. Zacharias notes that the Tycho-2 frame globally has the same accuracy as Hipparcos, however *not* locally due to the Astrographic Catalog plates and other early epoch ground-based positions: if one would have to average over many Tycho-2 star positions in say a 1 by 1 degree field, a systematic error limit would be hit at much larger error than that of Hipparcos star positions.

# 4. Optical and radio astrometry

### 4.1 Optical imaging of radio sources

Optical imaging of radio sources with optical counterparts is considered. The maintenance of astrometric reference frames in the long term is vital for the astrophysical analysis of high-resolution images obtained in different wavelengths. The importance of the reference frame may be illustrated with the supernova SN1987A in LMC. The optical image of the supernova was at first seen far off the centre of the radio ring which had been resolved by VLBI, but this enigma was solved when the preliminary astrometry from Hipparcos was made available for the study by Reynolds et al. (1995). The images moved by about 0.5 seconds of arc to coincidence because an error of this size in the previous optical astrometric reference system could be eliminated.

Usually however, the superposition is done by means of symmetry in the images or by use of the point sources if they are available in both images. Two examples are: Mahony et al. (2011) where the typical offset between radio and optical position is 1-2 arcsec, and Smail et al. (1999) where the radio and optical/near-IR images are aligned to an rms accuracy of 0.4 arcsec. In both case is the astrometry only briefly discussed, morphology being the main issue. N. Zacharias notes that both the 1-2 arcsec and the RMS 0.4 arcsec investigations seem to be very poor; with UCAC4 data today an absolute radio-optical alignment on the 50 mas level should be possible, even with only a few V=16 mag stars and approaching 20 mas if sufficient high S/N UCAC4 stars can be used.

The use of absolute positions as in the case of SN1987A was an exception, according to Ron Ekers (CSIRO Fellow, Australia) and he added: *"The best situation is for stellar radio sources. They will align until you start reaching the stellar diameter. But stars are radio weak so only a few will have high quality astrometry. SKA will change the SNR situation but SKA will have to include long baselines for astrometry. For Quasars it will depend on whether the c*

*orrect radio component is indentified with the optical AGN emission. This radio core will have a flat spectrum, or possibly synchrotron self absorbed due to its very small diameter. I would not trust any simple symmetry arguments as you normally have a core/jet radio structure and the question is how well can the core be recognised. When there are VLBI images with spectra I expect this is pretty good but as you start resolving the inner part of the radio jet (sub msec) it will break down. It's probably not known whether the optical AGN emission has offset structure. It would be assumed that this emission is from a very small diameter broad line region aligned with the radio core."* Zacharias notes that more important now seems to be the optical structure (unresolved) of host galaxies.

Ekers recalled lateron from his time as director of the VLA in the 1970s that he urged people to take the effort to use coordinates when superposing radio and optical images. Some did and some did not, but just shifted the images to a fit which sometimes resulted in wrong conclusions.

It is clear that the use of absolute positions in the alignment will become ever more important as the angular resolution and the available astrometric accuracy are improved. The asymmetry of some quasars in the optical was shown by Makarov et al. (2012) and this was discussed above.

The high angular resolution of present and future large optical telescopes will impose requirement of a milliarcsecond or less on the accuracy of optical astrometric reference frames, a requirement which can be satisfied by Gaia but only for some time. On a longer time scale a satisfactory reference frame can only be provided if a Gaia successor is launched in twenty years. This situation is discussed in the following.

### 4.2 Astrometric requirements in 50 years

The Gaia catalog will have a mean epoch about 2016 and contain a billion stars with G=6-20mag; the G magnitude is approximately the same as visual, V magnitude. Neglecting e.g. recognized binaries, a considerable fraction will constitute a Gaia reference frame. The accuracy of this frame in the future, e.g. in 50 years, is mainly determined by the accuracy of the proper motions. The expected standard errors of proper motions from Gaia are 0.0136 and 0.175mas/yr for respectively G=15 and 20mag at the mean epoch of 2016. This implies position errors in 2066 of 0.68 and 8.8mas from 50 years proper motion.

What is the ultimate accuracy of positions required by astronomers in 50 years? The requirement will be that images of fields obtained in different wavelengths can be superposed solely by means of the observed positions, without any hypothesis about the structures in the field. A reasonable hypothesis in case of SN1987A was that the optical supernova was at the centre of the radio ring, but to have it proven by means of accurate reference frames was very important. In case of a quasar or an active galaxy symmetry should not be taken for granted.

The example of superposition of observations of supernova SN1987A in LMC illustrates the great improvement in astrometry where errors up to 500mas could be found before Hipparcos provided a reference frame with an error about 1mas. But this accuracy was only obtained for the Hipparcos stars usually brighter than 8mag. The Tycho-2 Catalogue published in 2000 extended the frame to V=11.5mag with errors about 60mas. Zacharias note that UCAC4 gives positions accurate to about 20 mas on the ICRF down to R=14 mag today. Gaia will extend the frame to 20mag with errors about, e.g., 2mas for a time 10 years after the mid epoch of Gaia.

Consider an exposure at the wavelength of visual light $\lambda$=560 nm with a telescope of diameter D=50m for which the angular resolution is $\theta$=1.22$\lambda$/D=2.8mas. We can expect that diffraction limited performance with a 50m telescope will be obtained at some time in the future. The error of superposition n should be less than 0.1$\theta$=0.28mas and the error of a reference star position should be smaller. The adopted factor 0.1 is however rather arbitrary and we shall refer to it as the *modest requirement*.

Ideally, the error of superposition should be the same as the error of centroiding a point source, ie. $\theta$/sqrt(N) where N is the number of detected photons and the background is assumed to be negligible. With, e.g., N=10000 we would obtain a requirement of 0.01$\theta$=0.028mas, ten times smaller than the above "modest" value. The positional error expected with the 42 m aperture E-ELT is of similar size, 0.04mas according to Trippe et al. (2010) and the following discussion about the MICADO camera at the ELT. The aperture of the E-ELT has later been changed to 39 m which makes no difference for the present considerations.

Stars of G=15mag will have an error from Gaia of 0.14mas in 2026 which is barely satisfactory. It is however problematic to bridge the gap from G=15mag to the much fainter objects observed by the 50m telescope and there is no such bright star in the field of view. Therefore the faintest stars in the catalog must be used, i.e. G=20 which have errors of 1.8mas in 2026 which is far from good enough for the purpose, and in 2066 the error would be 8.8mas!

If a Gaia successor flies in twenty years the proper motions errors in a combined catalog will be ten times smaller and the position errors in 2066 will be 0.03 and 0.4mas, 20 times smaller because the epoch difference will be half as large. The errors quoted here for Gaia proper motions are minimum values since they do not take into account the distortion of proper motions from the numerous unknown binaries. Such errors are practically eliminated in the proper motions derived from two missions, secular proper motions are obtained.

In summary, a telescope with the diameter D=50m has a resolution of 2.8mas for $\lambda$=560 nm. It therefore needs an astrometric reference frame with accuracy 0.28mas or better at the time of observation in order to enable image superposition without any structure hypotheses. A Gaia successor flying in 20 years can provide a frame of G=20mag stars having sufficient accuracy even in 2066. With only one Gaia launched in 2013 the errors in 2066 will be at least 20 times too large even when comparing with the above defined *modest requirement*.

Richard Davies, Principal Investigator of the MICADO camera wrote: *"My thoughts about how soon we might realistically achieve the diffraction limit over a useful sized field on a 50-m class telescope are these: According to current shedules, the 30-40m class telescopes should be operational around 2024. This will provide diffraction limited performance at $\lambda$>1µm over a small 10-20" field very soon after. I expect that this will be extended to a 30-60arcsec field and decent sky coverage before 2030. To achieve similar at optical wavelengths is incredibly much harder (e.g. several orders of magnitude more computing power is required, even if one can make the necessary real-time measurements for the correction). Probably by 2030 we could do the diffraction limit on an 8-m at 560nm over a useful field (i.e. ~30") and with useful sky* *coverage (>10%) by 2030. For the ELTs, My guess is that would have to wait until at least 2050. But that matches the timescales you are talking about."*

### 4.3 Astrometry with VLBI

A correspondence with radio astronomers[2] in February and March 2014 is collected in Høg (2014a) from where I quote hereafter. All respondents agreed that the Gaia reference frame will not be sufficiently accurate in the long term to match the accuracy of the VLBI frame of the future.

Krichbaum wrote that todate the geodetic VLBI community is linking the radio and optical reference frame with an accuracy of the order of ~0.3 - 1 milli-arcsec. In 50 years the accuracy could be <= 10 micro-arcsecond when using the sub-mm band. At this level, however the internal source structure of quasars, etc. will be visible, which will make it hard to anchor a coordinate system. Zacharias notes that the optical - radio ref. frame link currently seems to be accurate only on the 3 mas level (Zacharias & Zacharias 2014); the quoted 10 uas is only for the radio reference frame, a link to the optical (Gaia) frame on that level is highly doubtful.

For comparison, linking of the Gaia reference frame to quasars should be inertiality as good as 0.5 µas yr$^{-1}$ according to Mignard (2011). This implies a deviation of 25 µas after 50 years for the link. I understand this to say that the linking of Gaia to quasars and to VLBI will be to the best standard of VLBI.

The situation is very different however when we consider the Gaia frame in a small area of the sky as required for use in a camera of a large telescope as discussed in this report for the MICADO camera of E-ELT. Only a few of the faintest stars at about 20th mag will be available for which the standard error of a position will be about 1.8 mas in 2026 and 8.8 mas in 2066. The latter would be improved to about 0.4mas by a Gaia successor mission and this seems to be the only realistic way to do so.

Krichbaum suggested that this problem of Gaia might be overcome by space-interferometers and others before him have thought the same. But this is now history. In retrospect, the great vision of optical interferometry from space, 1980-2010, may be called a dream, according to Høg (2014b) because all technical studies have been stopped and only one optical mission with interferometry has been realized, viz in the Hubble Fine Guidance Sensors.

Macquart wrote that if 200 microarcseconds is the projected benchmark in accuracy from Gaia for the next decade, it may well not be sufficient.

Godfrey answered to my specific question about the VLBI possibilities with the Square Kilometer Array (SKA) by giving two references. He continued to say that the astrometric capabilities with the SKA are aspirational - the design of the SKA is ongoing, and exactly what will be possible is not yet clear. It will require a significant commitment of observing time for the geodetic observing mode, and that will be competing for time with other science cases.

Godfrey drew attention to the opportunity to align multiple reference frames using millisecond pulsars in binary systems with a white dwarf companion. Precise position measurements of pulsars will allow the optical, solar system and quasar reference frames to be tied with great precision.

Further to this he added, GAIA will provide data on white dwarf companions to millisecond pulsars. SKA observations of the millisecond pulsars in such systems will provide interferometric and timing positions, thereby allowing the three reference frames to be tied together with similar precision.

---

[2] Ron Ekers, CSIRO Fellow, CSIRO Astronomy & Space Science, Australia; Thomas P. Krichbaum, Max-Planck-Institut für Radioastronomie, Bonn, Germany; Leith Godfrey, ASTRON, Netherlands Institute for Radio Astronomy, Postbus 2, 7990 AA, Dwingeloo, Netherlands; Laurent Loinard, Center for Radio Astronomy and Astrophysics, Morelia Campus of the National University of Mexico; Jean-Pierre Macquart, Curtin University, Perth, Western Australia; and John Reynolds, Australia Telescope National Facility, Epping, Australia.

Loinard explained that most of the people using the VLBA for astrometry (like himself) do small angle astrometry: They use a known astrometric source to register the observations, which is only (at most) a few degrees from the target. Depending on the source brightness, they get between 0.01 and 0.1 mas astrometry. Beyond 10 mu-as, they seem to bump into the current systematic limit. The astrometry does not get better with higher S/N. He believes that absolute astrometry (which he takes to mean wide field astrometry) can reach the same level of accuracy, but this would require dedicated (and quite time consuming) observations which have not yet been attempted.

Reynolds wrote that absolute positions from VLBI images are currently limited by the relative sparseness of primary reference sources and of time-dependent structure effects in individual sources at cm-wavelengths, at the several x 0.1mas level. Hence the emphasis on "densification" of the ICRF to the point where several reference sources are available in a given field, allowing internal motion to be more readily identified and allowed for. He does not believe tha t VLBI from space will give very accurate absolute astrometry.

Ekers added that the need for "densification" of the primary reference sources is coming about slowly with more observations using current technology but what is really needed is much greater sensitivity so weaker radio sources can be used. This is particularly true for the stars where we are currently limited to the most extreme radio emitting stars. Telescopes like SKA will increase sensitivity by 1-2 orders of magnitude so when VLBI capability is added they will dramatically change this scene. This will certainly all happen on a 50 year time scale.

My conclusion is that Gaia and a Gaia successor can provide a reference frame which in 2066 will have accuracy about 0.4mas per star over the sky and 0.2mas in any small area where at least 4 stars are available, which is 20 times smaller errors than with only one Gaia mission. The systematic accuracy would be a few tens of microarcseconds. This will not be quite as good as VLBI, but there seems to be no way to do better in the optical.

**4.4 Astrometry with the ELT**

A telescope with 50 meter aperture was taken above as an example for the far future, but already the European Extremely Large Telescope (E-ELT) with 42 m aperture will come with strong demands to an astrometric reference frame in the near future.

Observations with the coming E-ELT with D=42m should be taken as an example of how to superpose an optical image on a radio image for which the exact position is given. The size of field in the ELT camera MICADO (Trippe et al. 2010) is 53x53 square arcsec. The optical reference frame should therefore ideally contain several stars, say at least 10, within an area of 1 sq.arcmin around the object of study. The Gaia reference frame will contain nearly a billion stars, ie. 6.6 stars/sq.arcmin on average, but with large variations over the sky.

If the number of reference stars in the field is sufficient, a series of exposures is taken with the radio position at different places inside the field. Different exposure times are used in order to trace the possible optical features of the object. For all exposures the astrometric distortions are derived from the reference stars and based on laboratory calibrations of the camera.

If the number of reference stars in the field is small, perhaps as small as three stars, a linear transformation could be obtained if the stars are well spread over the field. This is important because the field rotation and the scale value are difficult issues according to Trippe et al. (2010). Additional exposures of a field nearby with many reference stars could be taken in order to derive the distortion parameters. To see how far these parameters are the same in neighboring fields requires separate investigation. Just one reference star could determine the zero point and that would be enough if orientation, scale values and distortions remain sufficiently constant, and that could be the case if this star lies very close to the radio source.

Ricard Davies wrote about exposures of a nearby field to derive distortion parameters: *"Although this is mentioned also in the Trippe paper, I am not yet convinced how useful it will be. The problem is that there are many issues related to adaptive optics and atmosphere that will inevitable be different between a science and calibration field, however much one tries to match them."*

Observations with the ELT camera could in principle be used to densify the billion-star catalog with fainter stars, ie. fainter than G=20mag around a given radio source. But this would require repeated observations of the field over several years in order to obtain parallaxes and proper motions as required to have a reference frame at the epoch of the radio image.

### 4.5 Absolute astrometry with the GMT

The Giant Magellan Telescope (GMT) with 25.7 m aperture located on the Las Campanas Peak in Chile should be mentioned here because it seems to offer a large field for astrometry of 2 arcmin diameter. Milton et al. (2003) show that such a large field is possible with the MCAO techniques for GMT discussed in the paper. If a field of this size would be available, 3 or 4 times larger than for the ELT, there would be sufficient Gaia reference stars over most of the sky to make absolute astrometry without use of galaxies as references.

Davin Malasarn (Director of External Affairs, Giant Magellan Telescope Organization) has informed me that "the GMT will likely do narrow angle astrometry (small fields of view) to look at things such as the orbit of stars around black holes at galactic centers. … we cannot be specific about the field size for astrometry at this point."

### 4.6 Absolute astrometry with the TMT

The Thirty Meter Telescope (TMT) with 30 m aperture on Hawaii will be able to provide high precision astrometry with the Wide-field InfraRed Camera (WIRC) in a field of 30 arcsec diameter, according to the TMT Instrumentation and Performance Handbook of 2010. This is half the diameter of the MICADO camera field, so that a significant number of Gaia reference stars for absolute astrometry will be found only in clusters and in the Milky Way. – All the three extremely large telescopes are expected to be observing from about 2020.

### 4.7 Astrometry with the MICADO camera – see Appendix A

I asked some questions about the paper by Trippe et al. (2010) to the authors and this led to a very fruitful correspondence with Richard Davies, the MICADO Principal Investigator in Garching. The following description of ELT astrometry represents the present views of Richard Davies (RD) as they have developed since the paper of 2010. The description given in Appendix A is much more detailed than I originally intended because Ric's prompt responses to my questions (14 emails from Ric in 10 days) gave me the opportunity to better understand the astrometric potentials of this magnificent telescope and camera.

### 4.8 Relative and absolute astrometry

For clarification follow here some conclusions, agreed with R. Davies. MICADO can obtain *relative proper motions* in a field thanks to the many stars. MICADO can obtain *absolute proper motions* in a field if it contains QSOs and distant compact galaxies or Gaia reference stars. MICADO can obtain *absolute positions and proper motions* but only if there are absolute reference points, ie Gaia stars in the field, and only this should be called *absolute astrometry*. For *absolute astrometry,* Gaia reference stars are required and the 1mas accuracy requirement in Section 4.10 of the paper applies. The 1mas requirement also applies for transforming the pixel scale to angular measure and deriving the orientation as needed for absolute proper motions.

Zacharias adds for explanation: "absolute" positions used to mean a coordinate system is derived from own observations (i.e. without any reference stars), like transit circle or maybe Hipparcos/Gaia (with global orientation = convention), while astrometry in a focal plane with reference stars is already "relative", although derived positions can be on an absolute (ICRF/Hipparcos) system. (My note here is that I follow the definition given in the introduction to "Science cases" where "absolute astrometry" means astrometric results in an inertial coordinate system, no matter how they are obtained.)

Davies agrees that an astrometry mission in twenty years time would be very desirable for providing new reference points for absolute positions for use with MICADO in the 20-50 years perspective. With only the first Gaia mission the positions of 20th mag stars degrade to unsatisfactory values of 3.5mas already in 2036, and 8.8mas in 2066.

The high-accuracy of Gaia positions is required to calibrate the scale value and orientation of the camera by observations of Gaia stars, but these values are not constant from one field to another and not even within a night, cf Section 4.3, especially because of the adapter-rotator on the telescope. The current plan by the MICADO team to calibrate the scale and orientation of the camera would be with, for example, HST fields where relative astrometry is known sufficiently well. Using fields with several Gaia stars would, of course, improve on this.

The absolute parallaxes obtained by reference to galaxies rest on the very safe assumption that these are very distant, that they have smaller parallaxes than 1 muas for the accuracies of measurement we have discussed above. The absolute proper motions by reference to galaxies rest similarly on the assumption that these have very small proper motions. If they are smaller than 1 muas/yr they would have negligible effect at the accuracies we have discussed. This proper motion value corresponds to a tangential velocity of 5 km/s at a distance of 1 Mpc. Since the reference galaxies will always have distances larger than 1000 Mpc no reasonable tangential velocity could affect the measurements discussed. Reference to galaxies is of course not possible in the zone of avoidance close to the galactic plane, but that region will be abundantly covered by Gaia reference stars.

**4.9 Relative astrometry with the ELT after a second Gaia mission**

If a Gaia successor is launched in twenty years a billion star reference frame will again be obtained of roughly the same stars as in the first Gaia catalog. The accuracy of the combined catalog will satisfy the assumptions of positions with sub-mas accuracy made by Trippe et al. during several decades after Gaia. This means that "astrometric accuracies of 40 µas for positions or better – with 40 µas/yr in proper motions - can be achieved in any one epoch of actual observations" (p.1126) if this combined catalog is used in a new reduction of old observations kept in the ESO archive for fields with a sufficient number of Gaia stars. – The value of 40 µas for the accuracy is quoted from the paper and maintained here while the slightly larger value 50 µas shown in the figure of Appendix A was used above as agreed with Davies.

We may furthermore assume that several new observations are obtained for the archive in the twenty year interval for these fields. Each of these will give the same accuracy and they can be combined. Just the first and last epochs give *absolute* positions with 40 µas err ors which combine to about 3 µas/yr. Combining all epochs, errors of 2 µas/yr should easily be obtained corresponding to about 1 km/s at 100 kpc distance. This result could be available shortly after the Gaia successor catalog is released in 2040, with the assumptions made here.

The assumption that 40 µas errors can be obtained for absolute positions would be true if the field contains so many Gaia stars that their positional errors per star between the two missions of about 250 µas for G=20mag combine to 40 µas. This requires 36 Gaia stars, but in practice 4 stars should be assumed so that 120 µas may be expected, giving a more realistic combined absolute proper motion accuracy of 6 µas/yr corresponding to 3 km/s at 100 kpc distance.

Can the same accuracy be obtained for *relative* proper motions and for the *absolute* proper motions by QSOs and compact galaxies by observations over a period of twenty years with MICADO, without any use of Gaia reference stars? Gaia reference stars and galaxies supplement each other since the first are only numerous at fairly low galactic latitudes where galaxies cannot be seen. This question is addressed on p.1127: "Proper motions of order 10 µas yr$^{-1}$ can be detected within few years of observations." At least 3 years of observations is required to distinguish proper motion from parallaxes which is always needed at the accuracies in question. The effect on proper motions from three years of observations due to orbital motions in unresolved binaries should be discussed and cannot be neglected in general. But such orbital motions can be detected by long series of observations.

The calibration issues are listed on p.1132, the last point being: "Additional regular dedicated on-sky calibration observations of sufficient astronomical targets, e.g. star clusters, as secondary tests." But this will not be possible without a Gaia successor mission since accurate reference point will not be available anymore already a few years after the present Gaia mission.

If the field contains a sufficient number of Gaia reference stars as, e.g., at low galactic latitudes, in cluster or galaxies there should be no problem of such calibration if data from two Gaia missions are obtained. In other areas of the sky with low star density it might be possible to choose an object of study which happens to have an adequate number of reference stars around just by chance. Densification of the reference catalogue at least in some areas shall be considered in the following.

# 5. Densification and maintenance of reference frames

### 5.1 Densification from space of the reference catalogue

Densification of the Gaia catalog itself in the present is a possibility to be considered. Gaia is sometimes overloaded with data to be transmitted, at other times there is unused transmission capacity. Such times could be used with priority for transmission of data collected from sparsely populated areas of the sky, or perhaps generally from galactic latitudes higher than 45 degrees. From these areas data could be collected to a fainter limiting magnitude than the usual G=20mag and be marked for transmission when the capacity is available. If the limit were set to G=21mag in these areas the stars at the limit would not always be detected and measured. Assume that they received only 20 detections during the mission instead of the usual 70, they would still obtain a standard error for the mean position better than 1mas. This takes into account that the number of photons is 2.5 times less and the number of detections is 4 times less which means sqrt(2.5x4)=3 times larger errors than the 0.25mas expected at G=20mag which is better than 1mas. If this is done in both Gaia missions at 20 years interval the proper motions for such a star will have an error about 0.05mas/yr so that positions at e.g. the epoch 2046 would still be about 1mas, thus fulfilling the requirement by Trippe et al. for astrometry with the ELT.

The gain by this strategy is obviously not great and probably not worth the effort. The number of stars per sq.arcmin at b=45 deg is 1.8 and 2.8 at respectively V=20 and 21 according to Allen III, so only one extra star is added on average.

The detection limit is not sharp any way so that data for many stars fainter than G=20mag will be transmitted even when this present limit is applied. This would help only little to fill the empty areas in the sky.

## 5.2 Densification from the ground

Densification in some selected areas could be done from the ground. A denser frame of reference stars can be obtained by observations with the ELT using the classical overlap method developed for photographic astrometry many years ago. An area of 4 sq.arcmin will usually contain sufficient Gaia stars to derive a linear transformation. It contains on average 4 stars at b=60 deg and we must assume that the field distortions have been calibrated by observations of other fields with a large number of Gaia stars. If the area of interest is covered by e.g. ten exposures with partial overlap it is possible to derive positions of all stars in the area by block adjustment. This must be repeated during several years in order to derive also parallaxes and proper motions for all the stars. It should be noted that all these astrometric parameters are *absolute* parameters and not only relative because they are tied to ICRS through the Gaia stars in the area.

A possible densification towards much fainter stars of about r=27 mag is discussed in the following section.

## 5.3 Maintaining a reference frame from the ground

Is it possible to maintain the Gaia reference frame by observations of Gaia stars from the ground in the decades after Gaia? With what accuracy could it be done? With which instruments? In some selected areas? In most of the sky?

I hope for answer to these questions from Norbert Zacharias, Chief of Cataloging Division, U.S. Naval Observatory, Washington D.C., since he is one of the most qualified person I can think of in this matter and he has expressed an interest when I asked him some time ago.

*Norbert Zacharias:* Answers on 8 April 201 4, some follow here and some are given elsewhere in the text.

1) Radio - optical reference frame link (i.e. Gaia - ICRF). NZ refers to the paper by Zacharias & Zacharias (2014) which points out that there could be a serious problem because of source structures for absolute positions when using only about 600 QSOs, as currently envisioned for this link. The Gaia to ICRF link could be limited to about 0.5 mas globally (at mean Gaia epoch).

The accuracy of both the radio (VLBI) and the Gaia (optical) systems would be much better individually, but the alignment of the systems would be limited. Being able to use many radio stars in a future radio reference frame (SKA ?) will likely be the key for a better link, while a second Gaia mission will not be able to help in this respect.

Fortunately absolute proper motions and parallaxes of Gaia are not affected, if the instrumental Gaia system proves to be as rigid as expected.

2) Densification of optical reference frame. NZ agrees with Dave Monet's estimates regarding LSST and ground-based optical astrometry accuracy for fainter stars. The URAT (USNO Robotic Astrometric Telescope) ground-based observing program "phase 1", using the "redlens" astrograph has a huge field of view (FOV) of 28 sq.deg, enabling about 15 sky-overlaps per year; however, it is not as accurate as Gaia will be. It also does not go fainter than Gaia and thus its main purpose at this point is the observation of very bright stars, not covered by Gaia and their link to the Hipparcos / Gaia frame. (My note: Gaia covers bright stars to 6 and probably to 2 mag.)

However, the same focal plane of 4 big CCDs (each 10560 by 10560 pixels of 9 by 9 micrometer) could be used at a new 1-meter class telescope. Such an optical design was studied almost a decade ago, see for example Laux & Zacharias, Astron. Soc. of the Pacific Conf. Ser. Vol. 338, p.106 (2005). An all-sky survey with such an instrument, and concentrating on the fainter stars (utilizing Gaia reference stars) could easily cover the 15 to 22 mag range, thus utilizing the most accurate Gaia stars and provide improved reference star positions for LSST. Such a "URAT phase 2" instrument would have a scale of 0.5 arcsec/pixel, well suited for typical 1 arcsec seeing conditions, while still covering a 1.5 by 1.5 deg sky area with a single CCD, thus about 9 sq. deg. per exposure. A 10-fold overlap per year of an entire hemisphere could be done with a single such telescope (very cheap as compared to LSST).

Centroiding errors of single such exposures could be as small as 2.5 mas for well exposed stars, up to about 50 mas at the faint limit. The atmosphere would contribute about 10 mas RMS to this and final catalog positions could be about 2 mas, if systematic errors can be controlled to that level. Such a project would be beneficial particularly for the link of high-accuracy Gaia stars to LSST data, although averaging over many 18 to 20 mag Gaia star positions to provide LSST with a reference frame might be sufficient. This of course assumes that Gaia's limiting magnitude is at least 19th to provide enough overlap with LSST (saturation at 18). If this turns out to be a problem, e.g. due to the current stray light issue or detector radiation damage, a URAT-type program becomes critical for the urgently needed link between Gaia and LSST. (My note: Gaia's limit will be 20 mag.)

3) MICADO accuracy of positions. It appears to me that the positional precision discussed in this draft manuscript regarding the MICADO performance (appendix and figure) is based entirely on the fit precision of a point source. A 0.5% of pixel size position fit error is reasonable for sufficient S/N ratio. However, there is little discussion about the mapping quality of the instrument over the 53" by 53" field of view (FOV). Even if most (static) field distortions can be calibrated out, there is a large "leap of faith" from the mentioned 60 mas initial distortions to the 50 uas goal. How about distortions which change from exposure to exposure? The atmosphere alone should introduce larger than 50 uas effects. The 1.6m telescope at NOFS, when used for relative astrometry in parallax work has a typical 2 mas error floor for 10 min exposures in a few arcmin FOV. The ELT has a much larger aperture and a redder bandpass which will help, but going from 2 mas to 50 uas is a factor of 40 - how is that possible?

How about a tilt of the MICADO focal plane? I believe any AO system compensates for some tilt of the incoming wavefront. Is there also a random focal plane tilt which would need to be determined with reference stars for each exposure beyond the linear "plate model"? In that case at least 4 or better 5 reference stars, well spread out over the focal plane would be needed for an astrometric reduction, plus the error propagation for targets somewhere in the FOV will be much larger than for a linear model. How good are the astrometric mapping properties of current large telescope AO systems?

4) General notes. Although a second Gaia mission in about 20 years is highly desirable for astrometry, as you outline in this draft, I doubt that it will be of high enough priority to get funded. For example a while back Platais et al. (2006, PASP 118, 107) began observations of selected astrometric calibration fields with 4-meter telescopes. There were good reasons for this project, e.g. to allow better astrometry of other projects, but not enough observing time was granted, so the urgently needed 2nd epoch data was never obtained. The alignment of optical to radio data on the sub-mas level is desirable, but what is the impact to astrophysics if we do not get this? Can certain assumptions substitute for the lack of this knowledge? How much is good enough? Who is willing to spend so much money on a somewhat improved radio to optical reference frame link with no significant other results which could drive a 2nd Gaia mission? (My note: there are many other results to drive a 2nd mission.) Most of the science of e.g. the upcoming large telescopes can be done with absolute parallaxes, absolute proper motions (using backgound, extragalactic sources) and relative and/or moderately accurate positions.

What else could a Gaia successor mission accomplish? With another mission in about 20 years we could "nail" the proper motions of stars very well (order of magnitude better than with a single mission). That would allow to separate out potential orbital motions much better than with a single mission. This could have a big impact on detecting exoplanets. A study should be performed along these lines to see what benefit a second Gaia mission provides over a single mission. Similarly the impact on mass determinations of binaries should be investigated. Are there options to go deeper with a Gaia successor mission? Is slower rotation possible, leading to longer integration times? Is a stare-mode mission competetive?

**Dave Monet:** From correspondence (in Monet 2014) with Dave Monet, Chief Scientist for Astrometry, U.S. Naval Observatory, Flagstaff, I have gained the impression that the LSST project holds the best promiss within a twenty year horizon or before to provide an accurate reference frame to about r=27th mag and fainter for about one half of the sky, Dec= -65 to +5 deg. It is much more in the Southern sky than the Northern sky. The telescope will be at Cerro Pachon which is in the AURA area that includes CTIO. Observations should begin by 2020 when data would be arriving at 10 Tbyte per night.

According to the Science Requirements Document (SRD) by Ivezic et al. (2011) it does *not* belong to the specified scientific tasks of LSST to establish or improve the reference system, but we might think in that direction although it would clearly be a major undertaking and it would require additional funding.

According to the SRD (p.23ff), a ten year mission is planned with about 1000 visits of all parts of the sky on the program. For the relative astrometry: *"The required median astrometric precision is driven by the desire to achieve a proper motion accuracy of 0.2 mas/yr and parallax accuracy of 1.0 mas over the course of the survey. These two requirements correspond to relative astrometric precision for a single image of 10 mas (per coordinate)."*

Dave Monet wrote that he based the astrometric numbers in the SRD on observational results from the work at USNO as well as on data collected from the Subaru Suprime public archive as well as special collections made with Gemini South and the CTIO 4-m. He expects that the proper motion accuracy will get down to 0.1 mas/yr, better than the number in the SRD.

The absolute astrometry specification is: *"The LSST astrometric system must transform to an external system (e.g. ICRF extension) with the median accuracy of Table 20."* This means 50 mas as design specification and 20 mas as "stretch goal". *"The delivered absolute astrometric accuracy may be fundamentally limited by the accuracy of astrometric standard catalogs."* I understand these values to mean absolute accuracy per visit. Monet explained that the 50 mas goal was set long ago (2005) when Gaia was more of a dream than a reality. Hence, he chose numbers that could be met with UCAC or similar reference catalogs. Monet is looking much forward to use the Gaia catalog and to cooperate with Gaia teams.

With 1000 visits over the ten year survey an accuracy for the mean position of 1.0 mas should be obtained since this is the expected accuracy for parallaxes. If the Gaia catalog is used as reference frame all these accuracies would then be the accuracies for *absolute* positions, proper motions and parallaxes if there are a sufficient number of accurate Gaia stars in the field and if they have the highest positional accuracy. The most accurate field in the LSST has an area about 20x20 arcmin$^2$ which would contain 2700 Gaia stars on average and at the galactic pole ten times less. The accuracy of Gaia positions for the G=20 mag stars will be 1.8 mas in 2026, ten years after the Gaia mean epoch, about the epoch for LSST and this accuracy is quite adequate. The resulting LSST reference frame could contain all stars to about r=27 mag with the mentioned accuracies for absolute astrometry. The resulting absolute proper motion accuracy of 0.2 or perhaps 0.1 mas/yr limits the 1.0 mas position accuracy of the frame to less than 10 years. But Monet has some worries, e.g. because the cadence of LSST covers the sky very quickly, and it is quite possible that the camera is never in true thermal equilibrium because the elevation and rotation will change every minute or so.

We shall now consider if the LSST or a similar instrument could produce a 1.0 mas reference frame about the year 2066 based on the Gaia reference frame. The Gaia positions for G=20 mag stars will in 2066 have an accuracy of 8.8 mas which would perhaps not be quite sufficient even in the presence of the large number of reference stars, e.g. 270 stars at the galactic poles. A remedy would be to use only somewhat brighter Gaia stars with better accuracy, but the LSST's sensors will saturate at something like 18th magnitude so that only the faintest Gaia star can be used, according to Monet.

My provisional conclusion is that an LSST instrument could maintain an astrometric reference frame with 1.0 mas positional accuracy during the coming 50 years when based on the Gaia frame. These observations would have to be continued almost uninterrupted during the 50 year period. The resulting LSST frame with stars to r=27 mag would contain a sufficient number of stars even for the small field of MICADO. The 1.0 mas accuracy is probably the utmost to be reached from the ground for faint objects and is therefore not matching the accuracy of a Gaia successor mission.

**Acknowledgements**

I am grateful for discussion and correspondence with Tansel Ak, Michael I. Andersen, Nils O. Andersen, Jeff Andrews, Frederic Arenou, Jean-Eudes Arlot, Coryn Bailer-Jones, Carine Babusiaux, Selçuk Bilir, Anthony Brown, Jos de Bruijne, Andrew Cooper, Poul Darnell, Richard Davies, Ron Ekers, Claus Fabricius, William Folkner, Jesús Zavala Franco, Paulo Freire, Johan Fynbo, Leith Godfrey, Andrew Gould, Hugh Harris, Amina Helmi, David Hobbs, Carme Jordi, John Leif Jørgensen, George H. Kaplan, Jun Yi Kevin Koay, Jean Kovalevsky, Thomas Krichbaum, Laurent Loinard, Jean-Pierre Macquart, Valeri V. Makarov, Davin Malasarn, Dave Monet, Alcione Mora, Max Palmer, Michael Perryman, John Reynolds, Samir Salim, Michael Shao, Matthias Steinmetz, Thomas Tauris, Sascha Trippe, Mattia Vaccari, Fred Vrba, Darach Watson, Christian Westhues, Norbert Zacharias, and Hans Zinnecker.

1204-1214. http://adsabs.harvard.edu/abs/2004MNRAS.351.1204V

Valenti, E.; Ferraro, F. R.; Origlia, L. 2004b, Red giant branch in near-infrared colour-magnitude diagrams - II. The luminosity of the bump and the tip. Monthly Notices of the Royal Astronomical Society, Volume 354, Issue 3, pp. 815-820. http://adsabs.harvard.edu/abs/2004MNRAS.354..815V

Zacharias, N.; Zacharias, M.I. 2014, Radio-Optical Reference Frame Link Using the U.S. Naval Observatory Astrograph and Deep CCD Imaging. The Astronomical Journal, Volume 147, Issue 5, article id. 95, 18 pp. (2014). http://esoads.eso.org/abs/2014AJ....147...95Z

van Langevelde H.J. 2012, The future of VLBI. Proceedings of Science: Resolving the Sky - Radio Interferometry: Past, Present and Future, Manchester, UK, April 17-20, 2012.
# 7. Appendix A

## Astrometry with the MICADO camera

**7.1  Overview**

I asked some questions about the paper by Trippe et al. (2010) to the authors and this led to a very fruitful correspondence with Richard Davies, MICADO Principal Investigator in Garching. The following description of ELT astrometry represents the present views of Richard Davies (RD) as they have developed since the paper of 2010. The description is much more detailed than I originally intended because Ric's prompt responses to my questions (14 emails from Ric in 10 days) gave me the opportunity to better understand the astrometric potentials of this magnificent telescope and camera.

The paper by Trippe et al. (2010) discusses in great detail relative and absolute astrometry in a field of 53"x53" with the NIR imager MICADO for the E-ELT. The 42-m aperture telescope will achieve diffraction-limited resolution of about 10mas at wavelengths of 2 μm. (The standard formula gives a resolution of $\theta=1.22\lambda/D=12$mas at 2 μm.) The instrument is sensitive to the wavelength range 0.8-2.5 μm, thus covering the *I, Y, J, H, K* bands, but not the visual band assumed in the above D=50 m example.

The expected precision is shown as function of magnitude in the figure. *Precision* is sometimes called *accuracy* in the paper, but it means the expected standard error per star from a one hour exposure. A*ccuracy* usually means the standard error of measurement in comparison with other independent and more accurate measurements and it is often larger than the *precision* because some error sources have not been recognized. The figure assumes 1 hour integration time, at least 30 minutes integration time per epoch is required in order to average out atmospheric tilt jitter.

The paper states on p.1139, III: When using high-z galaxies as astrometric reference points, integration times up to 10h can be necessary. This is explained by RD: "*In some cases "non-resolved galactic star clusters can be used as point-like reference sources ... by galactic star clusters we mean star clusters in distant galaxies - because at 12mas resolution even compact galaxies cannot be assumed to be point like, but will break up into discrete structures. Given the brightness of these star clusters in faint galaxies that we know, one needs 10 hours to get good signal-to-noise if one wants to use them as astrometric references. This is a worse case scenario since may galaxies will have bright compact nuclei which will do instead, or QSOs would also do - it's just a matter of being prepared for what might or might not be in any given field.*" The 10 hours integration can be distributed on several nights if needed, e.g. one hour per night of a field near the meridian as required especially for parallax measurements as proposed below.

The precision value 50 μas for stars brighter than K_AB=25mag is derived as 0.5 per cent of the resolution element since this is regularly reached at present 8-10 m class telescopes. The fall-off at fainter magnitudes than 25 is explained by RD as due to background: *"In the near infrared, and especially in K-band, for long exposures (assuming that each individiual integration within that long exposure is not read-noise* *limited) the noise is dominated by the background."* RD continues:

> *"For the photon rate of a K_AB=26 star I use the following: K_Vega = K_AB-1.85*
> *K_0_Vega = 4.66e9 ph/s/m$^2$/μm*
> *Ks bandwidth is 0.32μm telescope*
> *area is 1200m$^2$*
> *throughput is 0.8 * 0.8 * 0.5 (tel * AO * instrument)*
> *(where tel stands for telescope transmissivity, AO for Adaptive Optics system, instrument for MICADO camera)*
> *to get 125ph/s from the 26mag star.*
> *One hour in a source of K_AB=26mag would then give 4.5e5 detected photons from the source. The background would be around 6000 photons in a 3mas pixel; so within the 12mas FWHM of the PSF this would be 10$^5$ photons."*
>
> *"We read out several times and then add afterwards. Typically, one would read out as soon as the background (rather than readnoise) limit is reached. In a broadband filter this may take just a few seconds."*

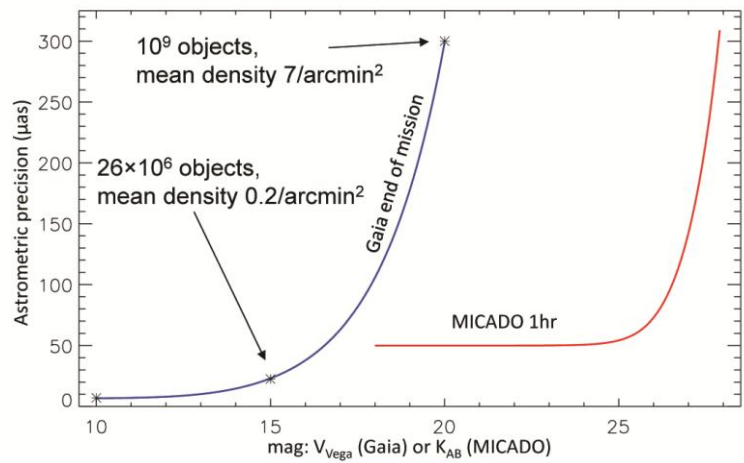

**Figure 4.** V$_{\_vega}$ refers to V-band magnitude in the Vega scale (i.e. A-stars are roughly zero magnitude in each band). The K$_{\_AB}$ refers to K-band magnitude in the AB system (in which the same magnitude in any band refers to the same flux density). Since there is ~2mag difference between K$_{\_vega}$ and K$_{\_AB}$ for the same flux density, and the colour of a 'typical' (e.g. K0) star is V-K=2mag, then I have suggested that for typical stars V$_{\_vega}$ ~ K$_{\_AB}$ and hence plotted them on the same scale. – Text and figure are courtesy Richard Davies.

The figure shows that Gaia and MICADO will obtain similar high astrometric precision, unmatched by any other optical astrometric instruments and similar to the highest astrometric precision obtained by VLBI. Gaia and MICADO aim for "very different science" as the figure says, but we shall point at possible interactions of the two and some are already mentioned in the paper.

The simplest application is to use Gaia absolute astrometry for calibration of MICADO exposures. This means to use the faintest stars in the Gaia catalog at 20th mag in order to have a sufficient number in the small field of view. Absolute positions are important when images of the optical counterpart of a radio source shall be superposed with the radio image obtained by VLBI. I asked if this task has been considered and RD answered: *"We had not addressed this during the Phase A concept study, since we had focussed on measuring the relative motions of objects detectable within the MICADO field over multiple epochs. However, this is an important point you raise, about being able to astrometrically match single epoch observations to data at other wavelengths. We will definitely add this item to our Phase B design study."*

A limitation on the direct use of Gaia reference stars is however their scarcity although there is one billion of them in the sky. "At least three reference sources are required for a full linear transformation" says the paper, but this number of three will not be obtained at high galactic latitude. At galactic latitude 60 deg there are on average 1.1 star per sq.arcmin brighter than V=20, according to Allen III, p.244. At b=30 and 45 deg there are respectively 3.5 and 1.8. The mean number given by Allen is 8.8, which is quite realistic compared with the 6.7 if there are a billion stars in the Gaia catalog. Three stars are required to derive the linear transformations, and the stars must be well spread over the field, in order to cope with the adapter-rotator instabilities which can introduce position variations across the field of 1 arcmin of the order 60mas, according to Section 4.3.

The MICADO team foresees extensive laboratory calibrations of the camera and expects to reach the 50 µas precision without the use of Gaia data. The team focuses on *relative astrometry* of the thousands of objects in the field, ie stars, compact galaxies and QSOs. This will provide accurate *relative proper motions* of the individual objects but not expansion, contraction or rotation in the field.

*Relative parallaxes* could be obtained but this is of little value unless very distant objects are present in the field. In fact there are distant objects. To the question about how many there are, RD answered: *"Deep number counts suggest there are at least 100 objects per sq.arcmin at K_AB~25mag, and maybe many more (our 5sigma detection limit in 1 hr is ~K_AB~28mag) - e.g. see figure 4 at Ferguson et al. (2000). How many of these will be good to use as astrometry references is not clear, because we don't know enough detail about galaxy structure on the relevant scales; but certainly we expect many to contain 1 or 2 bright unresolved sources within them (this is based on structure revealed by gravitational lensing of distant galaxies, etc). So to be conservative, let's say ~25 per sq.arcmin are good as references. This is uncertain, and is again something we need to look at more carefully in Phase B."*

*Absolute proper motions*, ie. proper motions in an inertial system can be obtained by use of the galaxies as references to tie the proper motions from repeated exposures of a given field together. The standard errors of the motions would be composed of the error of any given star and the systematic error of observing the galaxies in the field. This latter error would be 10 µas, assuming there are typically 25 galaxies each observed with an error of 50 µas in each exposure. Taking for example two exposures of a field at an interval of five years the standard error of the proper motions would be 10sqrt(2)/5=2.8 µas/yr for the systematic part. The errors of the individual motions are 50/5=10 µas/yr. The systematic error of 2.8 must be added quadratically and is therefore negligible.

### 7.2 Absolute parallaxes with the ELT

The MICADO team has focused on the measurement of proper motions and paid little attention to parallaxes, according to Richard Davies. He has therefore encouraged me to write in detail about parallaxes which I will do.

We adopt the conventional terminology in astrometry where the motion of a single star on the sky is composed of two parts: a straight *proper motion* due to the motion in space of the star and the solar system barycenter and a *parallactic motion* which is cyclic with a period of one year. For a binary star comes an additional motion due to the orbit. The parallactic motion forms an ellipse on the sky with its main axis equal to twice the parallax and parallel to the ecliptic.

The parallactic shift has its maxima during the year when the star is seen 90 degrees from the sun when the shift is equal to the parallax. The traditional method to measure parallaxes from the ground has been to observe the shift of the star relative to a frame of several stars in the field, preferably when the shift is near maximum. The difference of two such measurements with an interval of 6 months would give the relative parallax of the star if the proper motion is known. Therefore observations are obtained during several years from which the relative parallax and relative proper motion are derived by a least-squares solution.

Dave Monet was asked about the measurement of parallaxes described in the following four paragraphs and he answered in three mails, collected in Monet (2014).

All observations should be obtained close to the meridian in order to keep the variations of the instrument during the seasons and years to a minimum and to avoid the variations of the instrument from pointing it in different directions. The atmospheric effects are minimal in the meridian because the air mass is minimum and the effects are much same at all exposures. This means that all parallax observations are obtained during the early evening shortly after sunset and during the morning before sunrise. This practice is of course also recommended for the ELT. There could however be reasons to observe during the whole visibility season, but still only in the meridian, because binary nature of the stars could be better discovered.

It was common practice in photographic astrometry of parallaxes to use only the measurements in RA direction to derive the parallax and not those in direction of declination. These latter are affected by atmospheric refraction and dispersion and the parallactic amplitude in declination is much smaller, for both reasons these measurements were only used as check on the RA measurements.

In case of ELT measuring relative to galaxies, the *absolute parallaxes and proper motions* of all stars in the field would be obtained. Let us consider six observations during three years and separated by half a year as proposed. They should be analyzed by least-squares, but we consider here for simplicity that the proper motion is derived separately and subtracted from all six observations. Take the average of the resulting positions for the three morning observations and subtract it from the average of the three evening observations. The difference divided by two is the parallax. The standard error of this absolute parallax for a star is 0.5x50sqrt(2)/sqrt(3)=20 µas since we can neglect the error from the 25 reference galaxies.

In case a cluster is observed with thousands of stars at the same distance, the error of the cluster parallax will be dominated by the measurement error on the 25 galaxies. This error contribution is 0.5x10sqrt(2)/sqrt(3)=4 µas, assuming that all other error sources can be neglected.

It is noted that the paper does not consider magnitude depending errors in the error budget (Sect.5). --- RD answered: "*This is correct. For the purposes of the paper, we assume we are not limited by signal-to-noise or crowding effects, and that one can measure and correct for detector non-linearity.*" There could be such errors when bridging large magnitude differences unless the sensor is completely linear and the location estimator is ideal. Can this be assumed? --- RD answered: "*We have not investigated the impact of this in detail. The near-infrared detectors we plan to use (HAWAII 4RG) are highly linear to a large well depth; and there are standard procedures for measuring and correcting non-linearities.*"

# 8. Appendix B: Naming the Gaia successor mission

The Gaia successor mission is presently called Gaia2, but another name ought to be found. It is proposed to give the new mission the name *Rømer* or *Roemer*. The name Roemer was given by the thirteen proposers in May 1993 to their concept of an astrometric all-sky mission for the Third Medium Size ESA Mission (M3). We proposed direct imaging on CCDs in scanning mode as finally adopted for the large Gaia mission in January 1998 after an interferometric option had been investigated.

Ole Rømer is the name of a scientist who deserves a satellite to be called after him. An astrometric satellite matches especially well since he invented the meridian circle, the main instrument for fundamental astrometry during several centuries. Rømer stayed in France 1672-1681, he travelled to Holland and England and he maintained a lifelong correspondence with leading scientists in Europe. Rømer used his observations of stars, planets, the Sun and the Moon for many purposes, e.g. to determine the parallax of stars, the equinoxes, the precession and the length of the siderial year. All the while he served the Danish King in many duties, judge in High Court, chief of the police, Rector of the University etc. The following is expanded from the text in the M3 proposal.

The Danish astronomer *Ole Rømer* (1644-1710) is best known as the discoverer of the speed of light. Around 1675, while working at the newly created Royal Observatory in Paris, he noticed that the intervals between successive eclipses of Jupiter's moons were not always in agreement with the ephemerides that had recently been calculated by Cassini. Depending on the relative motion of Earth and Jupiter the intervals were sometimes larger, sometimes smaller. Rømer correctly inferred that these discrepancies were due to the finite time it took for light to travel from Jupiter to Earth. He computed a value of 22 minutes for the time it takes light to travel one diameter of the Earth orbit.

After his return to Copenhagen in 1681, Rømer constructed a series of instruments for measuring the positions of celestial bodies. His instruments gradually incorporated several new and ingenious concepts which were perfected during the next two centuries: the use of a long axis resting on two bearings for better definition of the viewing plane; microscopic reading of a graduated full circle; the use of counterweights to reduce flexure; observations in upper and lower culmination which made the instrument self-calibrating; and an emphasis on symmetrical design and measuring principles to eliminate otherwise uncontrollable errors. His *rota meridiana* constructed in 1704 is the prototype for the modern meridian circle, one of the most efficient and accurate instruments for ground-based positional measurements.

Rømer's strive for ever improved accuracy was strongly motivated by the search for stellar parallax. He made systematic observations of Vega and Sirius around the equinoxes and was quite certain to have found their collective parallax from right ascensions obtained with the transit instrument since 1692, as he wrote in a letter to G.W. Leibnitz dated 21 April 1703, but remaining doubt kept him from publishing the result. This phenomenon, the ultimate proof of the Copernican theory, would however elude astronomers for yet another century.

The following illustrates Rømer's continued efforts to improve the accuracy of his instruments. Observation of right ascension depends critically on the accuracy of the clock used in timing the star transits. In 1703 after 20 years of use, Rømer was able to correct his three clocks to within 0.5 second in 24 hours and his right ascensions were good within 7 arcsec, as claimed in the letter to Leibnitz. The rate of pendulum clocks changes with temperature because the length of the pendulum changes. Rømer had used a thermometer in 1692 to study the thermal expansion of metal and the effect on the clocks and the declination scale. He built thermometers using alcohol as the liquid in a glass tube. He used two fix points, freezing and boiling of water, he thus constructed and used the first calibrated temperature scale. In 1708 the 22 year old D.G. Fahrenheit visited Rømer and Fahrenheit mentions in a letter dated 17 April 1729 that be borrowed the idea for his temperature scale from this visit. All his happened long before G. Graham in 1721 developed the first self-regulating pendulum.

All Rømer's instruments and all the observations except those from three nights called *triduum* were destroyed in the fire of 1728; there were three volumes alone from the meridian circle. The *triduum* observations of 88 stars were used in 1756 by Tobias Mayer together with his own observations to discover that a fourth of the stars showed a significant change of position, thus deriving the first proper motions of stars from modern observations.

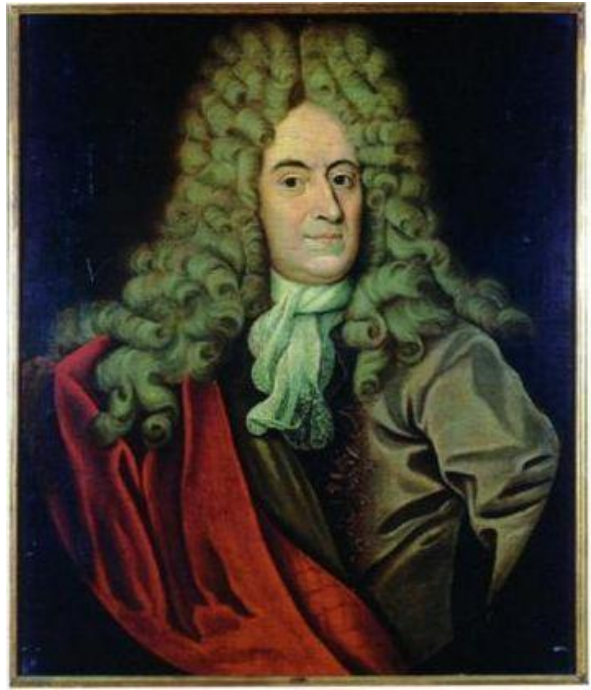

**Figure 5.** Ole Rømer (1644-1710).
Portrait by an unknown contemporary artist, exhibited in the Round Tower of Copenhagen.

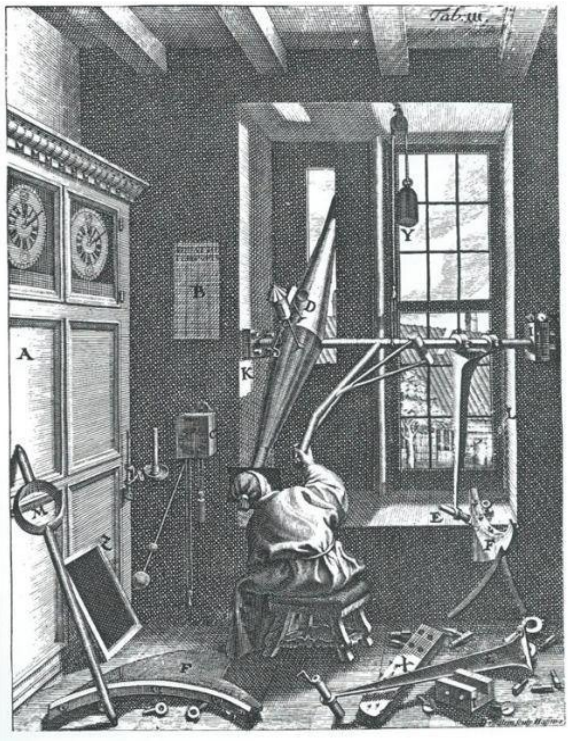 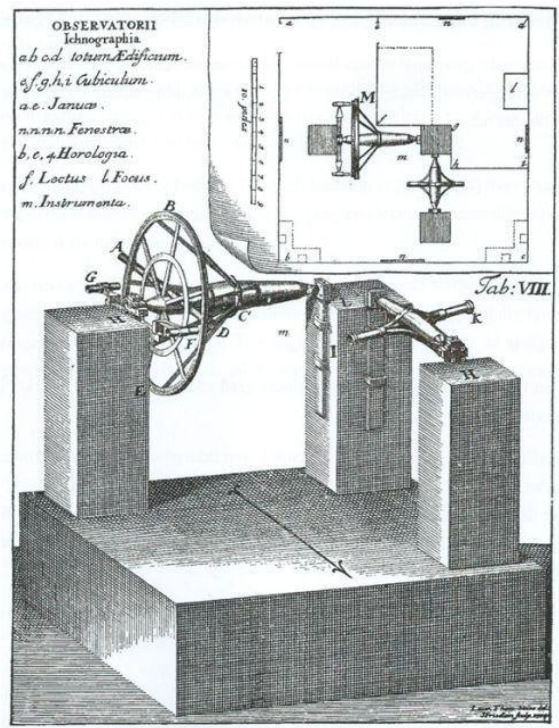

**Figures 6 and 7.** Ole Rømer's transit instrument from 1691 at left, which was used for observations from 1692.
The meridian circle *rota meridiana* from 1704 at right.
Illustrations from Peter Horrebow's *Basis astronomiæ* 1735.